\documentclass[10pt,
superscriptaddress,
showpacs,preprintnumbers,
 amsmath,amssymb,
prb,
floatfix,
lengthcheck,%
]{revtex4-2}
\usepackage{graphicx}
\usepackage{epstopdf} 
\usepackage{amsmath}
\usepackage{amssymb}
\usepackage{color}
\usepackage{dcolumn}
\usepackage{bm}
\usepackage{natbib}
\usepackage[normalem]{ulem}

\begin{document}
\title{Emergent grain boundary phases in stressed polycrystalline thin films}
\author{Mengyuan Wang}
\author{Ruilin Yan}
\author{Xiao Han}
\author{Hailong Wang}
\email{hailwang@ustc.edu.cn}
\affiliation{CAS Key Laboratory of Mechanical Behavior and Design of Materials, Department of Modern
Mechanics, CAS Center for Excellence in Complex System Mechanics, University of Science and
Technology of China, Hefei, Anhui 230027, China}
\author{Moneesh Upmanyu}
\email{mupmanyu@northeastern.edu}
\affiliation{Group for Simulation and Theory of Atomic-Scale Material Phenomena (stAMP), Department of Mechanical and Industrial Engineering, Northeastern University, Boston, Massachusetts 02115, USA}

\begin{abstract}
The grain boundary (GB) microstructure influences and is influenced by the development of residual stresses during synthesis of polycrystalline thin films. Recent studies have shown that the frustration between the preferred growth direction and rotations of abutting crystals to local cusps in GB energies leads to internal stresses localized within nanoscopic surface layers around the valleys and ridges that form at emergent boundaries (eGBs). Using a combination of continuum frameworks, numerical analyses and all-atom simulations of bicrystal $\langle 111\rangle$ copper films, we show that eGBs tune their surface morphology and rotation extent in response to external strains. Compression favors rotation to and growth of low energy GB phases (complexions) at eGB valleys while tension favors the transitions at eGB ridges, a reflection of the stress-induced mass efflux/influx that changes the energetic balance between interfacial and deformation energies. Molecular dynamics simulations of strained and growing bicrystal films reveal that the eGB phase transition is coupled to island formation at the surface triple junctions, providing a direct link between eGB phases and surface step flow. The interplay between eGB structure, morphology and mechanics emerges as a crucial ingredient for predictive understanding of stress and morphological evolution during film growth, with broad implications for multifunctional response of polycrystalline surfaces in a diverse range of surface phenomena such as surface mediated deformation, interfacial embrittlement, thermal grooving, stress corrosion, surface catalysis and topological conduction.\\  

\end{abstract}

\maketitle

\section*{Introduction}
Thin film performance is directly impacted by the residual stresses that arise during synthesis. These stresses induce and accelerate structural failure mechanisms such as cracking, delamination and thermomechanical fatigue, and are one of the leading causes of failure in thin film devices and related architectures~\cite{book:LewisAnderson:1978, tsf:EvansHutchinson:1995}. The stresses also impact thin film processing. The kinetic processes in play during film deposition are sensitive to residual stress distributions and their evolution, and this { interplay} can have a decisive effect on film quality~\cite{book:FreundSuresh:2004, tsf:Spaepen:2000}. Functional properties such as thermal~\cite{tc:BorcaTasciucDresselhaus:2000, tc:LeeHaque:2010, tc:LiSalje:2014} and electrical conductivity~\cite{elecgb:MayadasShatzkes:1970, elecgb:Palasantzas:1998, elecgb:BarmakCoffey:2014, elecgb:CesarGallGuo:2014, elecgb:BisharaDehm:2021} that rely on phonon and electron transport are also affected by { these stresses}. The effect is usually detrimental, although in strain engineered systems the understanding of film stress evolution offers the possibility of synthesis of films with tunable properties~\cite{tsf:Sander:1999, tsf:StrasserNilsson:2010, tsf:NarayanachariRaghavan:2016, tsf:FluriLippert:2016, tsf:FluriLippert:2017}. 

In polycrystalline thin films, the stresses are modified by the presence of the grain boundaries (GBs) and their junctions. Predicting the stress state of as-grown films is challenging as the stress during film growth is continuously tuned by the evolving microstructure. For example, the formation of GBs between {coalescing island grains} reduces the surface energy yet it leads to tensile stresses~\cite{tsf:Hoffmann:1966, tsf:NixClemens:1999, tsf:FreundChason:2001}. The reduction in GB area during subsurface grain growth can also lead to tensile stresses as the excess volume of the lower density GBs is absorbed within the bulk of the film~\cite{tsf:Chaudhari:1972}. Compressive stresses arise at the surface due to the insertion of diffusing adatoms into the grain boundary~\cite{tsf:FloroChasonSrolovitz:2002, tsf:PaoChasonSrolovitz:2002}. The overall stress evolution is controlled by the competing kinetics underlying these stress generating mechanisms~\cite{tsf:LeibThompson:2009, tsf:YuThompson:2014, tsf:VascoPolop:2017, tsf:AbadiasMartinu:2018, tsf:RaoChason:2021}.     

The GBs can also change their character as {the interfacial microstructure} evolves during film growth~\cite{gbt:Cahn:1982, gbt:Shvindlerman:1985, gbt:Rottman:1988, gb:CahnTaylor:2004, gbt:TangCarter:2006, gbt:FrolovAstaMishin:2013, int:Rohrer:2016, gbt:MeinersLiebscher:2020}. For example, {the grains} can rotate to lower their energy~\cite{grot:Martin:1992, grot:HarrisKing:1998, gb:GutkinOvidko:2005, grot:Upmanyu:2006} or modify their dislocation content as they migrate in response to local sources and sinks for dislocations~\cite{gb:SrinivasanCahn:2002, gb:CahnMishin:2006}. The presence of free surfaces and size constraints during early stages of film growth can further modify GB structures~\cite{gb:WangYe:2018}. At the film surface, emergent grain boundaries (eGBs) can dissociate~\cite{tsf:RadeticDahmen:2002}, or { form valleys and ridges through rotation of the adjoining grains that render the films inherently rough~\cite{tsf:ZhangBoland:2017}. These GB structural transitions can modify the kinetics of film growth, yet our current understanding of these phenomena is largely decoupled from the stress state of the films}. 

In this study, we focus on the effect of film stresses on the structure and geometry of { emergent grain boundaries in copper films. We use an energetic analysis together with continuum-scale computations of copper bicrystals to determine the stability of strained eGBs and their surface morphology.} The continuum framework allows us to key identify interfacial and mechanical parameters related to eGB mechanics. The framework also serves as a basis for understanding the response of eGBs to film stresses using all-atom simulations (molecular dynamics, MD) which naturally incorporate the effect of these parameters. We conclude by discussing the implications of eGB mechanics and morphology on film growth and related phenomena such as {thermal grooving}, interfacial embrittlement, fatigue behavior and surface  catalysis.  

\section{Theory}
\subsection{Background}
\begin{figure*}[htp]
\includegraphics[width=1.8\columnwidth]{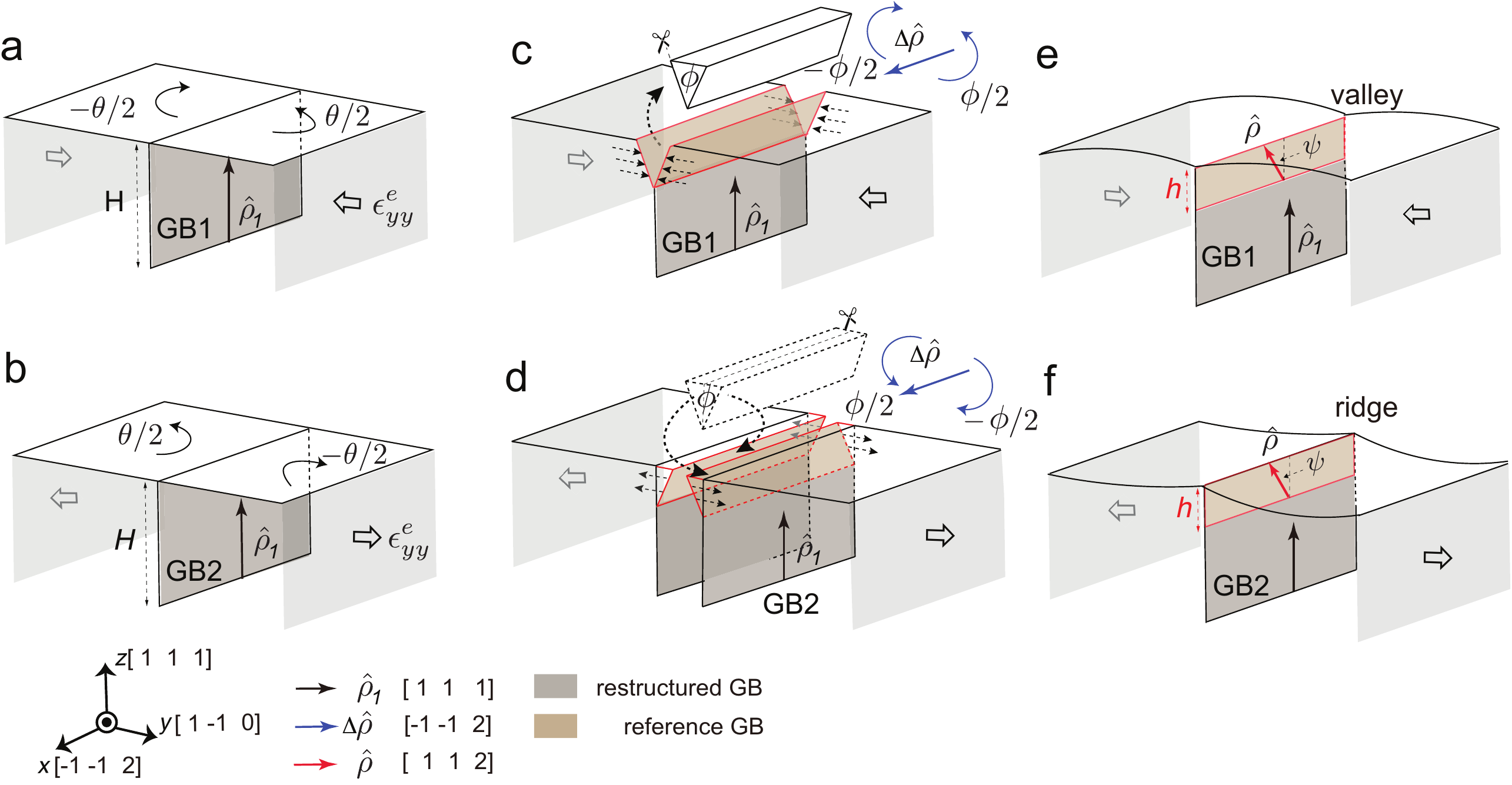}
\caption{{ (a-b) Schematic of an unrelaxed eGB at the surface of a thin subject to an external applied strain $\epsilon_{yy}^e$. The misorientation axis $\hat{\rho}$ of the symmetric tilt GB is along the film normal (arrow). The misorientation $\theta$ can be positive or negative depending on the relative rotation of the two grains.}  
(c) The cut-paste-shear stitch (CPSS) continuum framework used to analyze the stability of the rotated top layer at a valley formed by removal of a wedge of height $h$. The included angle of the wedge $\phi$ is the out-of-plane rotation associated with the $\langle111\rangle \rightarrow \langle112\rangle$ tilt of the GB misorientation axis.  (d) An analogous framework devised to study a ridge via {insertion} of a wedge at the eGB. See text for details. (e) The partial rotation of the grains within a top layer of thickness $h$ driven by the tilt $\psi$ of the misorientation axis within the GB plane (red arrow) to form a reconstructed eGB (solid red line) and an eGB valley. (f) A similar rotation results in formation of a ridge. The framework forms the basis for the energetic analysis of the restructured eGB within the partially rotated layer subject to film stresses. 
\label{fig:figure1}}
\end{figure*}

We begin our analysis by considering a generalized eGB bicrystal system composed of a flat GB {between} columnar grains terminating at the surface of a crystalline film, as shown in Figs.~\ref{fig:figure1}a and~\ref{fig:figure1}b. At low temperatures, thermal grooving at the eGB is negligible as the diffusion of surface steps is energetically unfavorable. Recent studies have shown that in nanocrystalline $\langle111\rangle$ copper films~\cite{tsf:ZhangBoland:2017}, symmetric tilt boundaries favor a tilt of their misorientation axis $\psi$ from the $\langle111\rangle$ {orientation set by the growth direction} to a neighboring $\langle112\rangle$ orientation. {For several GBs, the tilt of the misorientation axis within the GB plane permits the alignment of GB dislocations at the edges of the $\{111\}$ planes within the two grains. Due to the low stacking fault energy of Cu on $\{111\}$ planes, the dislocations can decompose into partials that lower the GB energy}. The geometrically necessary out-of-plane rotation of the abutting grains $\phi$ occurs about an axis at the intersection of the GB and surface plane, leading to formation of   valleys (Fig.~\ref{fig:figure1}c and~\ref{fig:figure1}e) or ridges (Fig.~\ref{fig:figure1}d and~\ref{fig:figure1}f). { The local angles confined to the eGB core, henceforth referred to as grooves, are consistent with atomic-scale structures of the restructure eGBs and the reconstructed surface TJs}.  Although { our understanding of these morphological changes at the eGBs is limited to nanocrystalline FCC films, the grain rotation that drives these changes} is expected to be more generally applicable in thin film systems where the (kinetically determined) growth direction~\cite{cg:Taylor:1992a, cg:DuSrolovitzMitchell:2005} is misaligned with respect to local cusps in bulk GB energies.     

{ When the film is constrained to grow along a preferred direction, say the $\langle111\rangle$ direction, the rotation of the grains cannot occur through the entirety of the film thickness. The rotation can abruptly terminate below a certain depth. Alternatively, the film can deform by developing elastic stresses in the vicinity of the eGBs which bring the film back into commensuration with the growth direction, thereby eliminating the need for the formation of the interface between the rotated layer and the remainder of the film. 
For a film of thickness $H$,  the reduction in GB energy (per film width) that drives the rotation scales as $\sim H$ while the elastic energy  scales with volume $\sim H^2$, indicating that that the elastic deformation itself is limited to a finite depth below the film surface.} 

Using a combination of atomistic and continuum simulations for both low-angle and high-angle GBs in $\langle111\rangle$ copper films, { Zhang {\it et al.}} have recently shown that the corresponding eGBs { restructure}, wherein the grain rotation is limited to a top layer of thickness $h\ll H$ {in the vicinity of the eGB core}, and each grain reorients back into commensuration with the film growth direction below this layer. The rotation is partial, confined to a wedge-shaped region around the eGB. The reorientation is realized through generation of elastic stresses that are also localized  to the restructured eGB. We use the terms `partially rotated layer" and `restructured layer" interchangeably to refer to the elastically stressed film layer containing the restructured eGB.

The scheme illustrated in Fig.~\ref{fig:figure1} serves as the basis for a { continuum-scale} computational framework employed to extract the thickness of the partially rotated layer. The morphological changes at the eGBs are induced by removing or adding wedges of height $h$ to form valleys or ridges, respectively. The included angle of these wedges is the local angle $\phi$ necessary for the tilt of the misorientation axis towards the $\langle112\rangle$ direction. The edges of the wedges correspond to the \{111\} planes at the atomic-scale. The modified eGB is then sheared so that the rotated layer does not form a new interface with the remainder of the film. The shear stresses are necessary to stitch the free edges associated with insertion or deletion of the wedges. At the atomic-scale, this corresponds to {an elastically stressed restructured layer} that is lattice-matched within the eGB plane as well as with the remainder of the film. Figures.~\ref{fig:figure1}c and ~\ref{fig:figure1}e show a  valley and a ridge formed using such a continuum-scale cut-paste-shear stitch (CPSS) scheme. The partially rotated state of the grains at the restructured eGB is consistent with scanning tunneling microscopy of the atomic-scale structure of the surface triple junctions (TJs) and the surface profiles~\cite{tsf:WangUpmanyuBoland}.

The surface morphology that accommodates the rotated top layer exhibits local grooves at valleys as well as ridges. The valley angles are driven by out-of-plane rotation $\phi$ and they slowly decay to zero at larger widths away from the eGB. The groove formation is not inconsistent with the interfacial force balance at the surface TJ - the elastic stress distribution can modify the interfacial free energies as well as the Herring torque terms associated with orientation dependence of the energies of the GB and that of the singular $\langle111\rangle$ surface. The corresponding elastic stress distribution (henceforth referred to as the intrinsic eGB stress) leads to co-existing GB phases - the rotated GB phase within the top layer and the unrotated (reference) GB phase below within the remainder of the film. Note that `GB phase'' is a misnomer here as these are two distinct boundary types with different macroscopic degrees of freedom. Rather, they are distinct interfacial states or complexions~\cite{gb:DillonCarterHarmer:2007, gb:CantwellHarmer:2014}. Here, however, the difference between the two GB structures is a local change in the orientation of the misorientation axis confined to a common GB orientation, giving the appearance of two GB phases with a well defined GB interphase line defect separating them for globally fixed macroscopic degrees of freedom. In the remainder of this article, we employ the term {`complexion" for each of the two GB types and use the phrase `eGB structural transition" to denote the local change in the eGB structure following grain rotation.}
\begin{figure}[htp]
\includegraphics[width=\columnwidth]{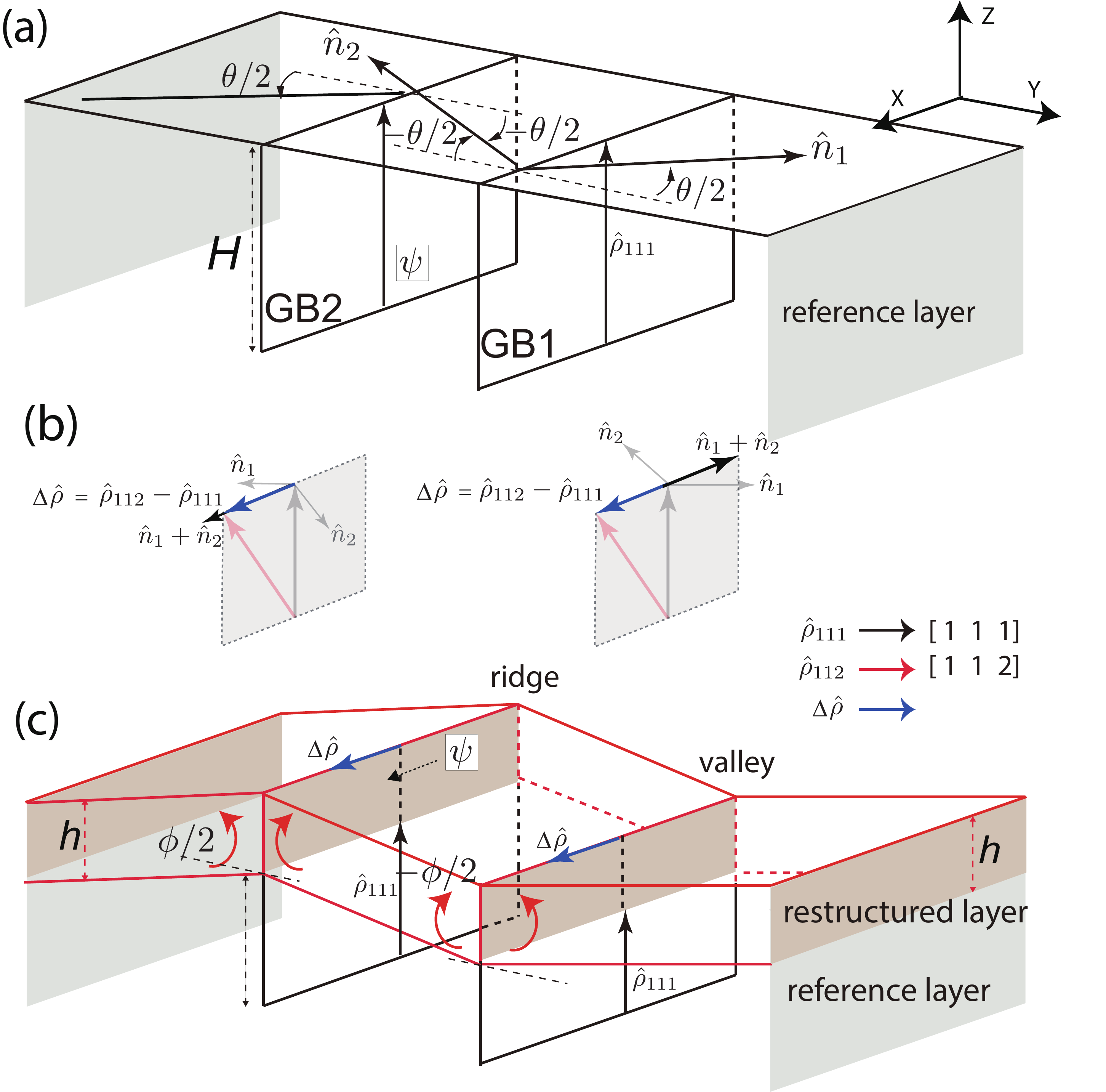}
\caption{{Schematic showing the structural transition of eGBs into valleys and ridges due to the tilt of the misorientation axis $\hat{\rho}$ to a neighboring GB energy cusp within a top layer of finite thickness $h$. (a) The reference GBs with $[111]$ misorientation axes terminating at the morphologically flat surface before the transition. (b) The sign of the eGB misorientation $\theta$ defined with respect to the vector along the change in the misorientation axis $\Delta \hat{\rho}= \hat{\rho}_{112}-\hat{\rho}_{111}$, and the crystal axis vectors emanating from the GB plane $n_1$ and $n_2$ (arrows),  as defined in the text. (bottom) The transition of the two eGBs to ridges and valleys. The sign of the out-of-plane rotation depends on the sense of the misorientation}.    
\label{fig:schematicValleyRidge}}
\end{figure}

\subsection{Energetic analysis of a strained eGB}
{The extrinsic film stresses and strains that arise during growth $([\sigma^e], [\epsilon^e])$ can influence the rotation-induced changes in the eGB structure as well as its surface morphology by modifying the intrinsic stress state of the eGB, $([\sigma], [\epsilon])$}. Figures~\ref{fig:figure1}c and~\ref{fig:figure1}d show the framework that forms the basis for an energetic analysis of the eGB stability within a stressed film. The cut$\rightarrow$paste$\rightarrow$shear stitch (CPSS) scheme lattice matches the rotated layer of thickness $h$ with the remainder of the film. For simplicity, we consider uniaxial deformation of the film. The extrinsic film stresses correspond to an effective (plane) {strain} normal to GB, $\epsilon_{yy}^e$. Then, the non-zero components of the extrinsic strain and stress tensors are 
\begin{align}
\label{eq:uniaxialPS}
& \sigma_{xx}^e=\frac{\nu E\epsilon_{yy}^e}{1-\nu^2}, \sigma_{yy}^e=\frac{E\epsilon_{yy}^e}{1-\nu^2} \nonumber \\
&\epsilon_{zz}^e=-\frac{\nu\epsilon_{yy}^e}{1-\nu}, 
\end{align}
where $E$ is the relevant Young's modulus and $\nu$ is the Poisson's ratio of the film.
   
Without loss of generality, we focus on a $\langle111\rangle$ symmetric tilt eGB at the surface of a $\langle111\rangle$ Cu film with overall thickness $H$ that undergoes a reorientation of the misorientation axis towards a neighboring $\langle112\rangle$ orientation. Based on the unrotated and stressed eGB as the reference (Fig. 1a), the total energy change ($\Delta U$) can be expressed as the sum of boundary ($\Delta U_{GB}$), surface ($\Delta U_s$) and elastic ($\Delta U_e$) energy changes,
\begin{align}
\label{eq:Energy1}
\Delta U (\phi, h, \epsilon_{yy}^e, H, L) = \Delta U_{GB} + \Delta U_{s} + \Delta U_{e},
\end{align}
where $L$ is the spacing between the eGBs or the average grain size at the film surface and $\phi$ is the out-of-plane rotation around the GB associated with the tilt of the misorientation axis $\psi$. There are two additional terms that scale with the length along the eGB, the energy of the surface triple junction and that of the interphase defect between the two {GB  complexions}. Since these are line defects their contributions are relatively small. This is consistent with previous atomistic studies on triple junction energies during grain rotation at the eGBs~\cite{tsf:WangUpmanyuBoland}. The change in TJ energies is negligible compared to the changes in the GB and surface energies.

{ The tilt in the misorientation axis and the out-of-plane rotation of the abutting grains $\phi$ are related geometrically}~\cite{tsf:ZhangBoland:2017},
\begin{equation}
\label{eq:phiCrystalGeometry}
\tan\frac{\phi}{2}=\sin\frac{\theta}{2}\tan\psi.
\end{equation}
Then, for a given tilt in the misorientation axis towards a local cusp, the eGB can form a valley or a ridge depending on the sense (sign) of the rotation angle $\phi$. {\it A priori} determination of the sign of $\phi$ requires the sign of the eGB misorientation as input, which we define with respect to a vector along the change in the GB misorientation axis $\Delta \hat{\rho}=(\hat{\rho}_{112}-\hat{\rho}_{111})$ that results from the added grain rotation at the eGB. Note that $\Delta \hat{\rho}$ is coincident with the periodicity vector associated with the atomic-scale structure of the surface TJ~\cite{tsf:ZhangBoland:2017}. Next, we define the crystal axes vectors $\hat{n}_1$ and $\hat{n}_2$ emanating from the GB plane. Within the perfect crystal $\theta=0$, these vectors are normal to the GB plane and antiparallel to each other. Following the in-plane rotation of the grains to form the GB with tilt misorientation $\theta$, the sum $\hat{n}_1 + \hat{n}_2$ is another vector along the surface TJ. For the flat eGBs prior to the formation of a valley or a ridge (Fig.~\ref{fig:schematicValleyRidge}a), the eGB misorientation is positive when the sum of the crystal axes vectors is parallel to $\Delta \hat{\rho}$, resulting in a valley (Fig.~\ref{fig:schematicValleyRidge}b and~\ref{fig:schematicValleyRidge}c). At a ridge, the two vectors are antiparallel and the misorientation of the eGB is defined as negative. That is,
\begin{align}
\Delta \hat{\rho}\cdot (\hat{n}_1 + \hat{n}_2) > 0 \rightarrow \theta > 0, \nonumber\\
\Delta \hat{\rho}\cdot (\hat{n}_1 + \hat{n}_2) < 0 \rightarrow \theta < 0. \nonumber
\end{align}
 Fig.~\ref{fig:schematicValleyRidge} schematically summarizes this interplay between bicrystallography and geometry at two eGBs that differ in the sign of their misorientation.

The two interfacial contributions  $\Delta U_{GB}$ and $\Delta U_{s}$ arise due to the changes in the GB and surface areas following the rotation. The elastic energy cost of the restructured layer has contributions from deformation of the bulk, the GB and the surface,
\begin{align}
\Delta U_e = \Delta U_{e,b} +\Delta U_{e, s} + \Delta U_{e,GB} \,.
\end{align}
Below, we develop expressions for each of these contributions within a strained film that serve as inputs for an energetic stability analysis of  the restructured eGB.  

\subsubsection{GB energy contribution}
The GB energy gain is the change in bulk energy of the GB within the { partially rotated} layer due to the $\langle111\rangle\rightarrow\langle112\rangle$ tilt of its misorientation axis,
\begin{align}
\label{eq:energyGB}
\Delta U_{GB} = h \left[\frac{\gamma(\phi)}{\cos(\frac{\phi}{2})} - \gamma_{\langle111\rangle}\right]\,,
\end{align}
where $\gamma(\phi)$ is the energy of rotated grain boundary. More generally, the misorientation axis searches for a local cusp in the vicinity of the growth direction. { For $\langle111\rangle$ symmetric tilt GBs in FCC crystals such as copper, the misorientation axis oriented along the $\langle112\rangle$ direction $\hat{\rho}_{112}$ is one such local cusp in the vicinity of the $\langle111\rangle$ preferred growth direction while preserving the GB inclination. The tilt allows the dissociation of GB dislocation partials on $\{111\}$ planes~\cite{gbe:SuttonBalluffi:1987, tsf:WangUpmanyuBoland}. Complete reorientation of the axis along the $\langle112\rangle$ direction results in a tilt in the misorientation axis $\psi=\psi_{112}$ and a geometrically related out-of-plane rotation $\phi=\phi_{112}$.}

\subsubsection{Surface energy contribution}
The rotation into valleys or ridges changes the total surface energy of the eGB system as the surface area changes. This is apparent within the CPSS scheme shown in Fig.~\ref{fig:figure1}c-d. {The surface area change is proportional to the area along the surface of the cut-out wedge with interior angle $\phi$},
\begin{align}
\label{eq:surfEnergy}
\Delta U_{s} =  \mp2\gamma_s \,h \tan\left(\frac{\phi}{2}\right)\,.
\end{align}
The change is positive and negative for valleys and ridges, respectively. Based on Eq.~\ref{eq:surfEnergy}, the surface energy reduction drives the formation of the valleys while it opposes the formation of a ridge.

\subsubsection{Elastic energy contribution}
{\it Bulk elastic deformation}: For small strains, the bulk elastic deformation is the net effect of rotation-induced intrinsic stresses and the extrinsic (film) stresses, expressed with respect to the {strained and unrotated bicrystal},
\begin{align}
\label{eq:bulkElasticEnergy1}
\Delta U_{e,b}  =  &  \,2\int_{-H}^0\int_0^L \left[(\sigma_{ij}+\sigma_{ij}^e)(\epsilon_{ij}+\epsilon_{ij}^e) \right]dydz \nonumber \\
& - 2\int_{-H}^0\int_0^L \sigma_{ij}^e\epsilon_{ij}^e\, dydz\,.
\end{align}
For uniaxial plane-strain deformation (Eq.~\ref{eq:uniaxialPS}), this simplifies to
\begin{align}
\label{eq:bulkElasticEnergy2}
 \Delta U_{e, b} = & \int_{-H}^0\int_0^\infty \left[\sigma_{ij}\epsilon_{ij} + \frac{2E \epsilon_{yy}^e}{1-\nu^2} \epsilon_{yy}\right] \,dydz \nonumber\\
= & f(\phi, h, H, L) + \epsilon_{yy}^e\, g(\phi, h, H, L)\,,
\end{align}
where the function $f$ is the contribution due to rotation-induced intrinsic stresses, and $g$ captures the interaction between intrinsic and extrinsic stresses. We note that the latter is unaffected by shear deformations.

{\it Surface elastic deformation}:
The elastic deformation of the $\{111\}$ surface is in response to { the strains $\epsilon_{yy}$ normal to the GB plane}. It can be expressed in terms of a set of generalized surface stress and surface elastic constants $\tau^0_{yy}$ and $S_{yy}$ confined to a surface layer~\cite{nano:Cammarata:1994, nw:MillerShenoy:2000}, 
\begin{align}
\label{eq:surfDeform}
\Delta U_{s, e} = & 2\int_0^L\left[\tau^0_{yy}(\epsilon_{yy}+\epsilon_{yy}^e)+\frac{1}{2}S_{yy}(\epsilon_{yy}+\epsilon_{yy}^e)^2\right] dy \nonumber\\
& - 2\int_0^L\left[\tau^0_{yy}\epsilon_{yy}^e+\frac{1}{2} S_{yy} (\epsilon_{yy}^e)^2\right] dy\,,\\
= & 2\int_0^L\left[\tau^0_{yy}\epsilon_{yy}+\frac{1}{2}S_{yy}(\epsilon_{yy})^2 + S_{yy}\epsilon_{yy}^e \epsilon_{yy} \right] dy\,. \nonumber
\end{align}
The response is again defined with respect to the reference unrotated and strained bicrystal, captured by the last term.

{\it GB elastic deformation}: For completeness, one can also consider the deformation of the GB in response to the film stresses. Similar to the surface layer, this effect involves interfacial stress/elastic constants for the boundary { complexions} before and after the rotation of the top layer,  
\begin{align}
\label{eq:elasticityGB}
\Delta U_{GB, e} = & \int_{-H}^{-h}(\tau_{\langle111\rangle}\epsilon_{zz}+\frac{1}{2}S_{\langle111\rangle} \epsilon_{zz}^2) dy  \nonumber\\
& + \int_{-h}^0(\tau_{\langle112\rangle}\epsilon_{zz}+\frac{1}{2}S_{\langle112\rangle} \epsilon_{zz}^2) dy\,,
\end{align}
where $\tau_{\langle111\rangle}/S_{\langle111\rangle}$ and $\tau_{\langle112\rangle}/S_{\langle112\rangle}$ are the effective constants. Given the ability of GBs to absorb stresses efficiently~\cite{gb:VitekSuttonSchwartz:1983, def:Swygenhoven:2003b, gb:HanVitekSrolovitz:2016}, we expect the { GB deformations} to be much smaller compared to the bulk and surface deformations. We therefore ignore the GB elastic response in the remainder of the article.

Our analysis identifies the interfacial and bulk parameters that serve as inputs to the GB, surface and elastic contributions. We quantify these parameters for pure copper using atomic-scale computations. Scaling analyses together with continuum computations of the elastic deformations are then employed to extract the dependence of the interfacial and elastic energy contributions on the film strain. Minimization of the total energy  $\Delta U$ for a specific eGB within a strained film (fixed $\phi$, $\epsilon^e_{yy}$, $H$ and $L$) yields the equilibrium height of the rotated layer $h^\ast$ that stabilizes the {restructured  GB  complexion} below the film surface. 

\begin{figure}[htp]
\includegraphics[width=\columnwidth]{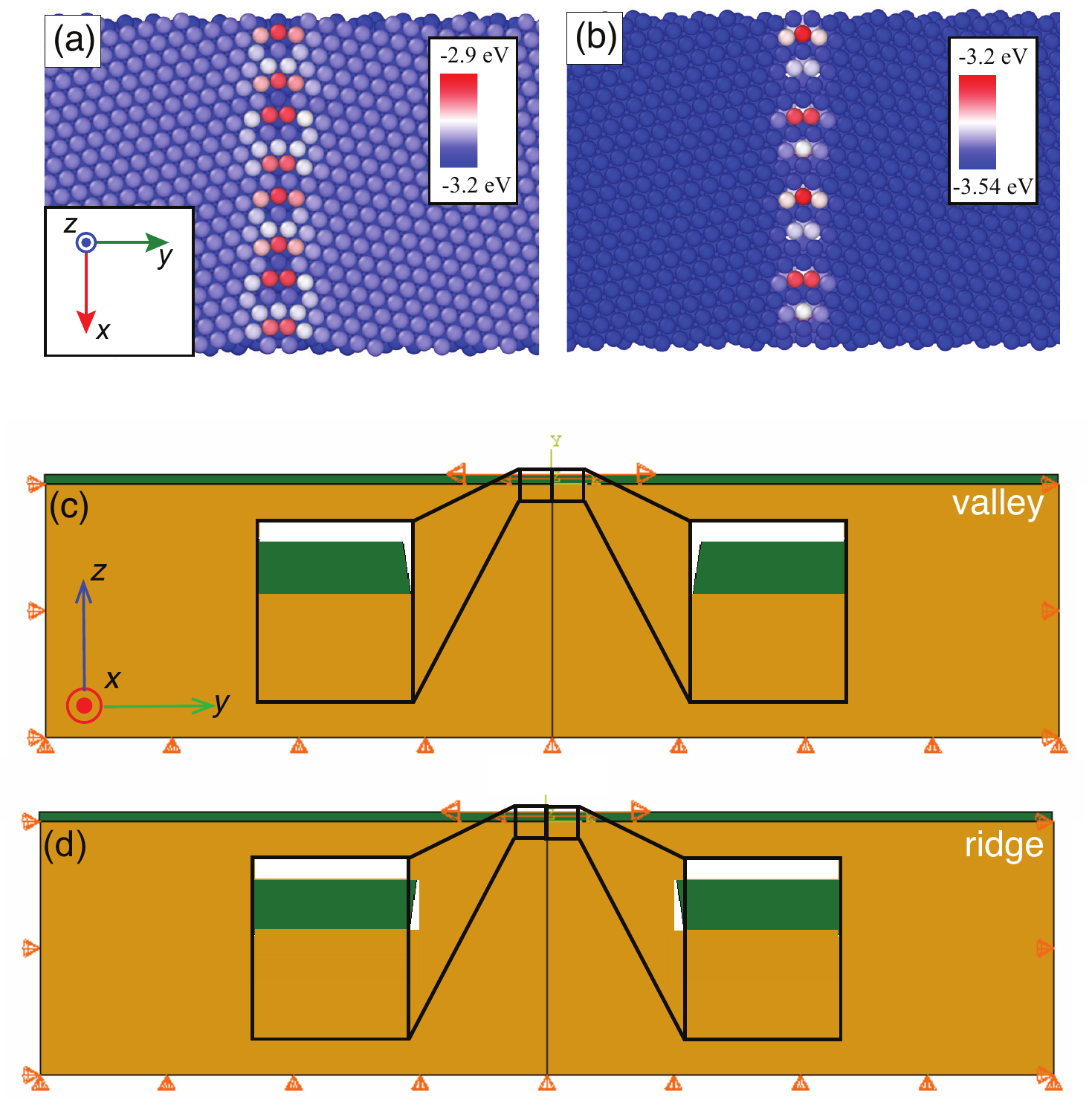}
\caption{(a) The atomistic computational cells used to study (a) eGBs in thin films (top view) and (b) bulk GBs (sectional view) in pure copper. The atomic configurations of the eGB and the bulk GB in (a) correspond to the $\theta=26.01^\circ$ high angle eGB (HAeGB) system in a $\langle111\rangle$ copper film. The color scale indicates the atomic potential energy.  (c-d) The configurations generated as part of the cut-paste-shear stitch (CPSS) scheme to calculate the elastic displacement fields around (c) a valley and (d) a ridge within the the FEM computations. The partially rotated top layer (shaded green) is shear stitched and its bottom surface is mated to the remainder of the film (shaded brown).
\label{fig:atFEMCompCells}}
\end{figure}
\section{Computational Methods}
\subsection{Atomic-scale simulations}
The atomic-scale computations are employed to: i) extract relevant interfacial and bulk parameters, ii) characterize the subsurface structure of the eGB bicrystals in response to external strains, and iii) study the mechanistic pathways for the structural transitions of the eGBs. The molecular statics and dynamics simulations are based on an embedded atom-method potential fit to equilibrium and non-equilibrium properties of copper~\cite{intpot:Mishin:2001}. Figure~\ref{fig:atFEMCompCells}a  shows the bicrystallography and atomic configuration of the {computational cell} used to study the $\theta=26.01^\circ$ eGB in a $\langle111\rangle$ copper film. The corresponding {computational cell} used to extract the equilibrium structure and energetics of the same GB in the bulk is shown alongside in Fig.~\ref{fig:atFEMCompCells}b. {It differs from the eGB computational cell (Fig.~\ref{fig:atFEMCompCells}a) in that it is periodic along all three directions with no free surfaces.}

The as-constructed eGB bicrystals 
consist of a GB terminating at the surface of $\langle111\rangle$ oriented Cu film. The eGB cell sizes are typically $10\,{\rm nm}\times 30\,{\rm nm} \times 10\,{\rm nm}$ ($250,000$ atoms in total). { Unlike the scheme shown in Fig.~\ref{fig:figure1}, the atomic-scale computations are performed at isolated eGBs by applying external strains or subjecting the surfaces to deposition fluxes. Larger configurations with thicker films ($>1$\,M atoms) are also simulated to eliminate size effects that can arise due to elastic interactions between the eGB and the edges of the computational cells}. The eGB cells are periodic along the $x$-direction with free surfaces along the $y$ and $z$ directions. Three layers on these free surfaces are fixed along their normal directions. The computations are performed using constant temperature within a canonical (NVT) ensemble. To study the effect of temperature, the simulations are performed at two temperatures, $T=700$\,K and $T= 1000$\,K. A Nos\'e-Hoover thermostat~\cite{atsim:Nose:1984} with fixed time step of $1$\,fs is employed to accurately capture diffusive events at the surface and along the GB~\cite{surf:WangShinzato:2021}. 

Finite temperature canonical (NVT) MD simulations are performed with a 1 fs time-step and a N\"{o}se-Hoover thermal bath. We use a genetic algorithm with varying atoms within the GB region to calculate both its ground state ($T=0$\,K) and finite temperature ($T=700$\,K) bulk structure and enthalpy. {The atomic density within the GB region is changed by inserting or deleting atoms within the GB core}. The computational framework involves generation of several trial configurations with varying atoms within a $10$\,nm wide region around the grain boundary core, and at each step the lowest energy configuration is chosen for subsequent trials. The energy of the computational cell is monitored until it reaches a steady-state and yields the equilibrium GB structure and enthalpy. 

Atomic-scale computations of eGBs subject to extrinsic strains are performed to study formation of the partially rotated layer, the structure of eGBs and the stress distributions associated with the co-existing GB { complexions}. The strain-free structures of both GBs serve as starting points for studying their response to extrinsic strains. The film is strained by changing the simulation box size along $y$ direction (Fig.~\ref{fig:atFEMCompCells}a) at a strain rate $10^6~\rm{s^{-1}}$. Lower strain rates are used to ensure that there is no rate-related artifacts. The strained configurations are then relaxed at the desired temperature using canonical MD simulations until the interaction energies converge.

Some of the structural transitions are facilitated by influx of atoms to the film surface. These simulations are performed by adding deposition clusters or nanometer-wide equilibrated monolayers all along the surface TJ. The configurations correspond to various stages of step-flow mediated thin film growth. We also perform thin film deposition simulations to study the formation and evolution of the structural transitions. These simulations involve three types of atoms within the computational cell: i) fixed atoms (bottom 6 layers, fixed), thermostated atoms (middle layers, canonical NVT ensemble), and Newtonian atoms (top 6 layers, constant energy NVE ensemble). The velocity of deposited atoms is $100$\,m/s directed vertically downward towards the surface. The simulations are performed for deposition rates in the range $0.1-1$\,atom/ps.  

The atomic distributions of the interaction energy, virial stress tensor~\cite{atmethods:VitekEgami:1987}, 3D central symmetry parameter~\cite{atmethods:KelchnerHamilton:1988} and dislocation analysis (DXA)~\cite{viz:Stukowski:2012} are used to dynamically characterize the structure of the bulk, the GB and the surface TJ. The finite temperature and strained configurations are relaxed at $T=0$\,K using molecular statics (MS) simulations for ease of visualization. These analyses are performed within the Open Visualization Tool (OVITO)~\cite{viz:Stukowski:2009}. The surface profiles are extracted by tracking the $z$-coordinates of the atoms at the surface. In several instances, the profiles of the subsurface are also monitored to extract the depth dependence of the surface profiles. 

The diffusion of the atoms within the GB { complexions} $D_{GB}$ is computed by monitoring the temporal evolution of the ensemble averaged mean square displacement (MSD) of the GB atoms, $\langle Z^2_{\rm GB}\rangle$. Linear fits based on the Einstein relation $D = (\langle Z^2_{\rm GB}\rangle)/2t$ are used to extract the diffusion coefficients. Each computation is performed over a $t=10$\,ns time interval and the MSD $\langle Z^2_{\rm GB}\rangle$ is extracted along the $\langle 111\rangle$ direction within a $5$\,nm wide window centered around the { GB core}.

\subsection{Continuum computations}
The bulk and surface elastic deformations of the eGB bicrystals subject to uniaxial strain are studied using a linear elastic constitutive law in a finite element method (FEM) package (ABAQUS 2017). The {computational cells} for valleys and ridges for a given out-of-plane rotation $\phi$ are shown in Fig.~\ref{fig:atFEMCompCells}. The bicrystal film is modelled using 4-node bilinear plane strain quadrilateral elements (CPS4R in ABAQUS notation). Grains of identical dimensions are used to create the computational cell, which within the FEM framework is essentially a single crystal that yields the (linear) elastic deformation associated with partial rotation of the top layer. Since the continuum computations do not explicitly model the structure of the GB or the thin film surface, they ignore modifications to the intrinsic structure of these interfaces and the GB { complexions} due to the external strain. 

The sides and bottom of the computational cell are fixed while the surface remains free. To avoid the influence of boundary conditions, we typically set the bottom and sides far from the eGB, $L/h=100$ and $H/h=50$. The plane-strain deformation yields the initial reference state (Fig.~\ref{fig:figure1}a). The CPSS scheme~\cite{tsf:WangUpmanyuBoland} using a notch of depth $h$ and interior angle $\phi$ is then employed within the strained cell to extract the sum total of the extrinsic and intrinsic elastic deformation of the {restructured eGB} (Figs.~\ref{fig:figure1}c and~\ref{fig:figure1}e). The stress distributions and elastic energies of the healed notch are monitored. The surface profiles are extracted using the nodal displacements.

\begin{figure}[htp]
\includegraphics[width=\columnwidth]{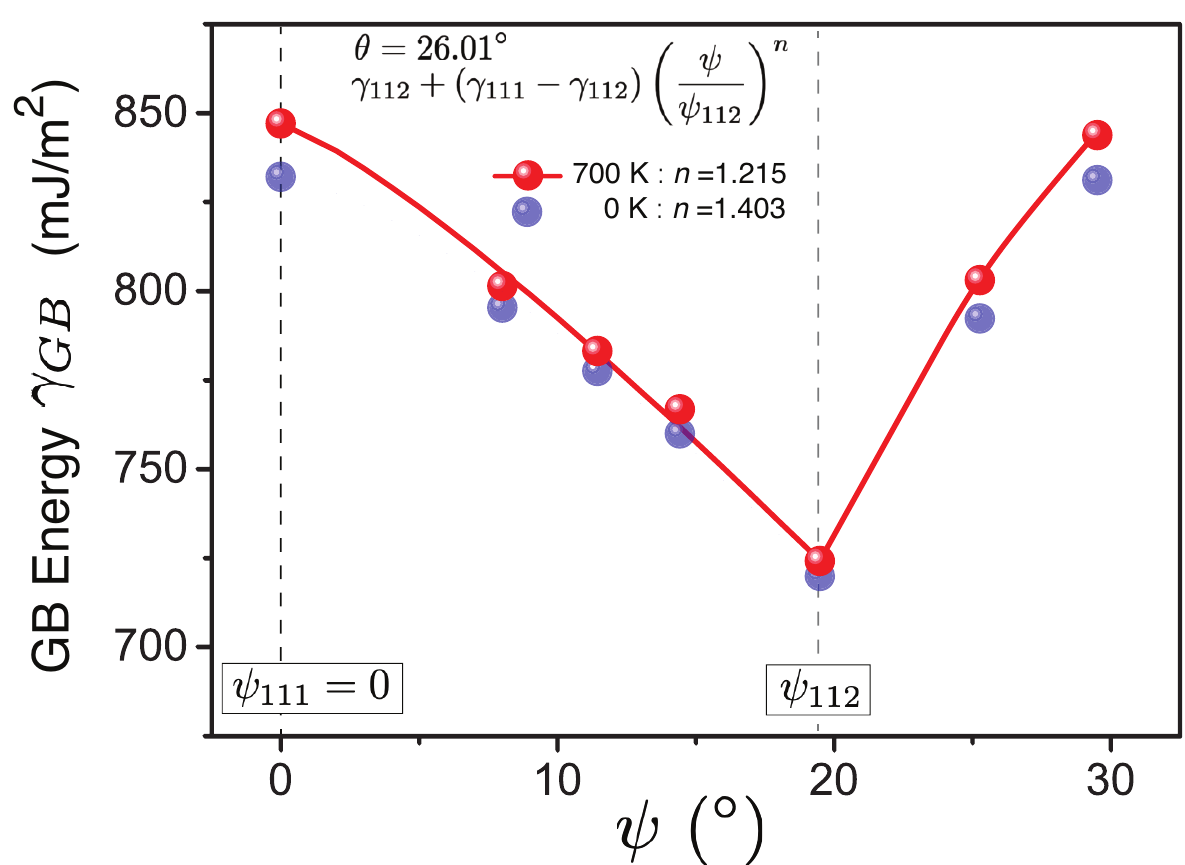}
\caption{The energy (enthalpy) $\gamma$ of the HAGB with $\theta=26.01^\circ$ in-plane misorientation at $T=700$\,K plotted as a function of misorientation axis tilt $\psi$ away from the $\langle111\rangle$ orientation. The energies at $T=0$\,K are also plotted for comparison.
\label{fig:GBEnergies}}
\end{figure}
\section{Results}
\subsection{GB energies, $\Delta U_{GB}(\phi)$}
Figure~\ref{fig:GBEnergies} shows the bulk GB energy (enthalpy) of the $\theta=26.01^\circ$ symmetric tilt HAGB at $T=700$K as a function of the $\langle111\rangle\rightarrow\langle112\rangle$ tilt of the misorientation axis $\psi$. The ground state energies extracted at $T=0$\,K are also shown for comparison; they are in agreement with prior efforts based on MS simulations~\cite{tsf:WangUpmanyuBoland}. The GB enthalpy decays monotonically towards a cusp corresponding to the $[112]$ axis. The effect of temperature is smaller at the $\langle111\rangle\rightarrow\langle112\rangle$ axis orientation and it increases away from this cusp. The energy variation is not symmetric about the cusp as the $\langle111\rangle$ orientations are not symmetrically distributed about the $\langle112\rangle$ orientations. The finite temperature energies require additional computations and we delegate them to a subsequent study. However, the enthalpy values and the nature of the structural transition allows us to estimate the GB free energy change associated with the rotation. Denoting $\gamma_{112}$ and $\gamma_{111}$ as the bulk GB energies corresponding to the two orientations of the misorientation axis, the enthalpy change $\gamma_{111}^H-\gamma_{112}^H$  is likely an upper bound as the finite temperature entropic reduction $-T\Delta S$ is usually larger for more disordered, higher energy GBs~\cite{gbe:SuttonBalluffi:1987}. 
\begin{figure*}[htp]
\includegraphics[width=1.8\columnwidth]{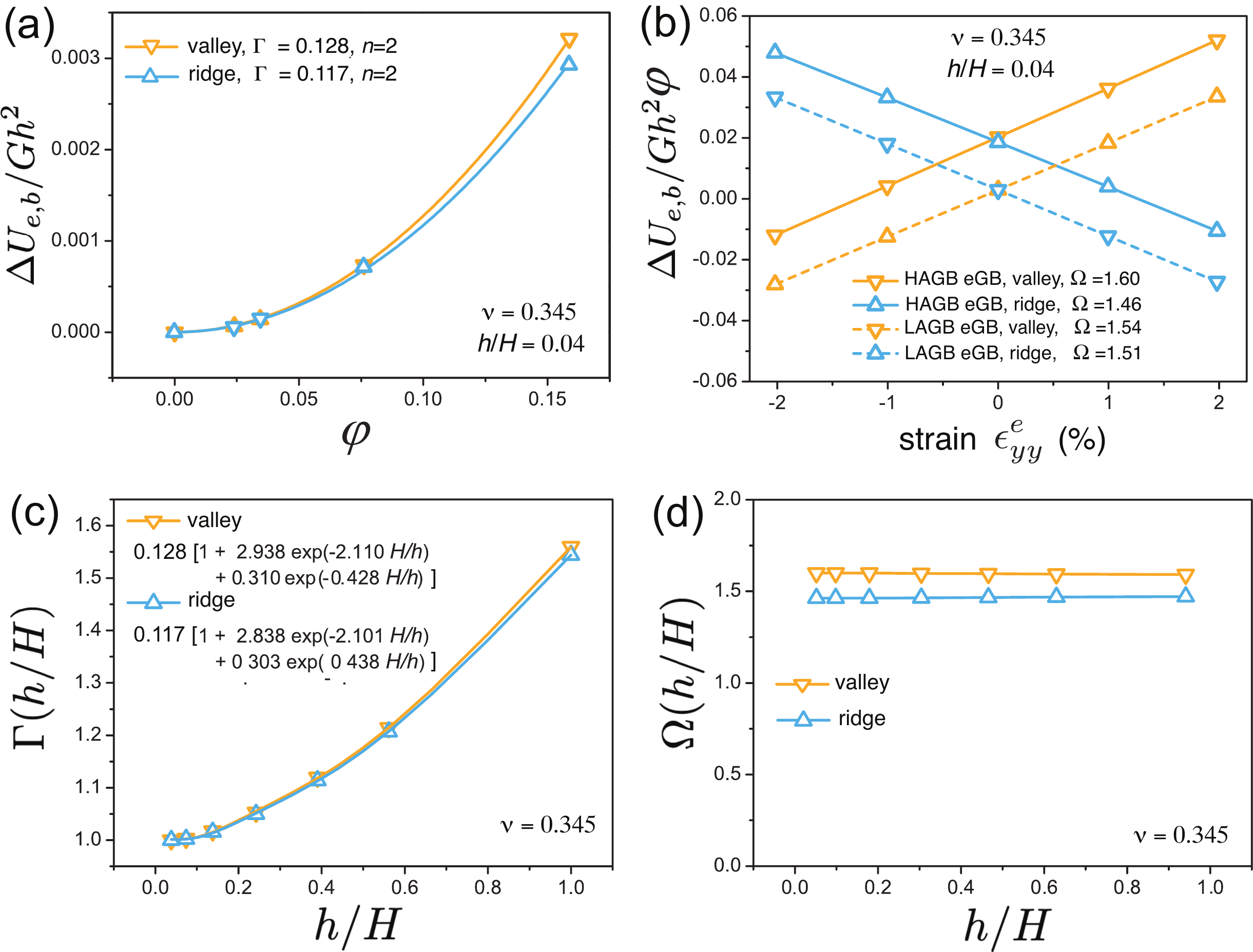}
\caption{(a) The elastic energy change $\Delta U_e/(G h^2)$ {associated with the formation of a valley and a ridge at a restructured eGB as a function of the out-of-plane rotation $\phi$}. (b) The elastic energy change $\Delta U_e/(G\phi h^2)$ as a function of external strain $\epsilon_y^e$ for the HAGB ($\theta=26.01^\circ, \phi_{112}=9.907^\circ$) and the LAGB ($\theta=3.89^\circ, \phi_{112}=1.375^\circ$) eGB systems. The size effect parameterized by (c) $\Gamma$ and (d) $\Omega$ as functions of the thickness ratio $h/H$ using FEM simulations for compressive strain $\epsilon_{yy}^e=-1\%$.
\label{fig:BulkElasticEnergyScaling}}
\end{figure*}

For the tilt angle $\psi\equiv\psi-\psi_{111}$ with respect to the unrotated $\langle111\rangle$ axis ($\psi_{111}=0)$, the decay in the GB energy can be approximated as 
\begin{equation}
\label{eq:GBFitting}
\gamma[\phi(\psi)]=\gamma_{112}+(\gamma_{111}-\gamma_{112}) \left(\frac{\psi}{\psi_{112}}\right)^n\,,
\end{equation}
with the power exponent $n$ the sole fitting parameter.
 Combining Eqs.~\ref{eq:phiCrystalGeometry} and Eq.~\ref{eq:GBFitting} allows us to express the GB energy as a function of the out-of-plane rotation, $\gamma(\phi)$. For an eGB within a $\langle111\rangle$ film with misorientation tilt axis between the $\langle111\rangle$ and the $\langle112\rangle$ orientations, $\gamma(\phi)$ and Eq.~\ref{eq:energyGB} together yield the contribution of GB energy change to the driving force { for the formation of the restructured eGB}.
It is a maximum for a complete reorientation of the GB misorientation axis towards the cusp, or for $\psi=\psi_{112}$ and $\phi=\phi_{112}$.

\subsection{Theoretical analyses and computations}
\subsubsection{Bulk elastic energy cost $\Delta U_{e, b}$}
For an infinite thick and wide film $h/H\ll1, h/L\ll1$, the deformation is determined by extrinsic stresses, material parameters and the geometry of the V-shaped notch that serves as the precursor {for the formation of the partially rotated layer}. A dimensional analysis shows that both the intrinsic deformation energy  as well as the interaction energy between extrinsic and intrinsic deformations vary quadratically with the rotated layer thickness, i.e. $f(\phi,h) \propto G h^2$, and $g(\phi,h) \propto G h^2$, where $G$ is the shear modulus. In particular, for a linear elastic material the elastic energy is proportional to the shear deformation, $f(\phi,h) \propto (\phi h)^2$ and $g(\phi,h) \propto (\phi h) h$. 
Then, for $h/L \ll 1$, the bulk elastic energy based on Eq.~\ref{eq:bulkElasticEnergy2} can be expressed as 
\begin{align}
\label{eq:functionsf-gScalings}
U_{e,b} = \Gamma\left(\frac{h}{H}\right)\,G\,\phi^2 h^2 + \epsilon_{yy}^e \,\Omega\left(\frac{h}{H}\right)\,G\phi h^2\,,
\end{align}
where $\Gamma$ and $\Omega$ are size corrections due to the finite thickness of the film.

The bulk elastic energy calculated using FEM computations of shear stitched notches is plotted in Fig.~\ref{fig:BulkElasticEnergyScaling}. The Poisson's ratio is taken to be $\nu = 0.345$ for pure copper. Varying $\phi$ and $h$ in the absence of external strain ($g=0$) yields a non-linear dependence of the elastic energy on the shear strain $\phi h$. This represents the intrinsic stress field generated by the {grain rotation}, captured by the function $f$. A quadratic fit yields the material constants $\Gamma(0)=0.128$ and $0.117$ for valleys and ridges, respectively (Fig.~\ref{fig:BulkElasticEnergyScaling}a). 

The effect of external strain captured by the function $g$ is shown in Fig.~\ref{fig:BulkElasticEnergyScaling}b. The energy is plotted for both valleys and ridges formed at the eGB. The variation for another eGB formed by the termination of a $\theta=3.89^\circ$ (out-of-plane rotation $\phi_{112}=1.375^\circ$) low angle GB (LAGB) is qualitatively similar. For linear elastic deformations, the energy varies linearly with external strain for varying thicknesses of the rotated layer $h$, in accord with Eq.~\ref{eq:functionsf-gScalings}. Curve fits yield the material constant $\Omega(0)$ and the values for valleys and ridges are indicated in Fig.~\ref{fig:BulkElasticEnergyScaling}b. For both eGBs, the elastic energy of ridge formation is lower under tension while compression strains favor the formation of valleys, indicating that this is a general trend for { restructured eGBs within partially rotated layers}.
\begin{figure*}[htp]
\includegraphics[width=1.7\columnwidth]{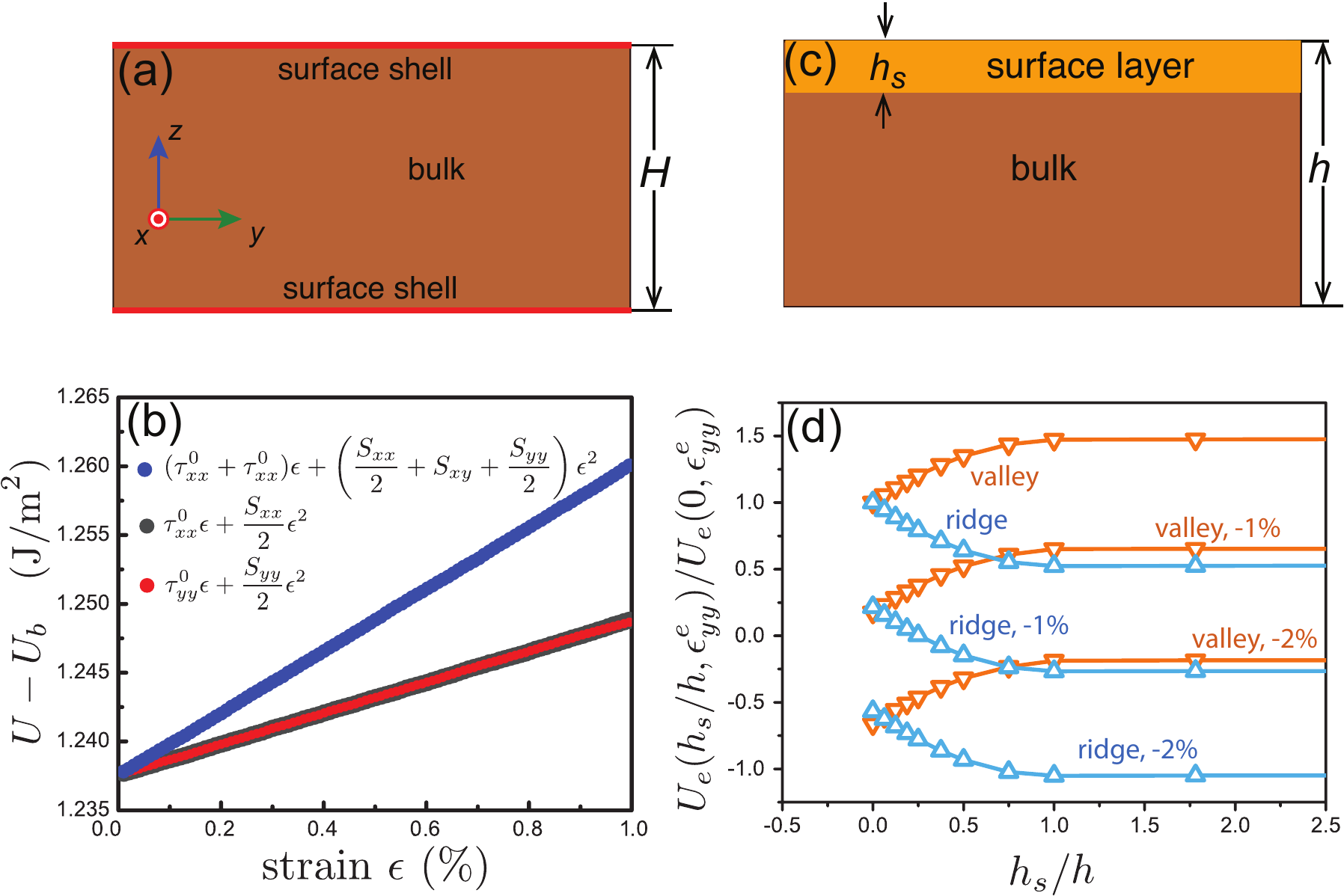}
\caption{(a) Schematic showing the bulk-surface composite framework composed of a copper bulk and two 2D surface shells of negligible thickness. (b) Surface deformation energy in unixial and biaxial atomic-scale simulations as a function of applied strain along the normal ($y$) and parallel ($x$) directions, ($\epsilon_{xx}$, $\epsilon_{yy}$). Curve fits yield the initial surface stresses ($\tau_{xx}^0$, $\tau_{yy}^0$) and the surface elastic constants ($S_{xx}$, $S_{xy}$, $S_{yy}$). (c) FEM computational cell with a finite surface thickness $h_s$ employed to study the effect of the ratio $h_s/h$. (d) The scaled elastic energy change $U_e(h_s/h, \epsilon_{yy}^e)/U_e(0, \epsilon_{yy}^e)$ of the valley and the ridge as a function of the (scaled) surface layer thickness $h_s/h$ for the HAeGB system.
 \label{fig:surfaceCompositeLayer}
}
\end{figure*}

Computations with varying $h$ at fixed film thickness $H$ allow us to quantify the finite size effect. The correction to the intrinsic elastic stresses $\Gamma (h/H)$ for the high angle eGB (HAeGB) system for rotated layer thicknesses in the range $0<h/H\le1$ is plotted  in Fig.~\ref{fig:BulkElasticEnergyScaling}c. The variation for valleys and ridges is similar and is well-described by a biexponential
\begin{align}
& \Gamma \left(\frac{h}{H}\right)=\Gamma (0)\left[1+A_1\exp \left(\frac{B_1 h}{H} \right) + A_2\exp \left(\frac{B_2 h}{H}\right)\right],\nonumber \\
& \Omega \left(\frac{h}{H}\right)=\Omega(0)\,.
\end{align}
The values of the set of constants $\{A_1, B_1, A_2, B_2\}$ obtained from fits to the FEM computations for the HAeGB are indicated in Fig.~\ref{fig:BulkElasticEnergyScaling}c. The functional form is simply a reflection of the exponential decay of the through-thickness strain field away from the stitched notch, modified by the finite thickness of the film. As the film thickness becomes smaller, the scale function $\Gamma$ increases with increasing $h/H$ as seen in Fig.~\ref{fig:BulkElasticEnergyScaling}c. Following Eq.~\ref{eq:functionsf-gScalings}, the trend indicates that the mechanical constraint imposed by the unrotated layer enhances the shear deformation in the rotated layer for both valleys and ridges, thereby decreasing $h^\ast$. Below a critical value of the film thickness, the rotated layer is unable to form. 

Figure~\ref{fig:BulkElasticEnergyScaling}d shows the variation $\Omega(h/H)$ at ridges and valleys within the HAeGB system for compressive strains.  Unlike the intrinsic function $\Gamma$, the interaction between the intrinsic boundary strains and and extrinsic applied strains is no longer mediated by shear deformation (second term in Eq.~\ref{eq:bulkElasticEnergy2}), so the scale function $\Omega$ is almost independent of the ratio $h/H$.
\begin{figure*}[htp]
\includegraphics[width=1.8\columnwidth]{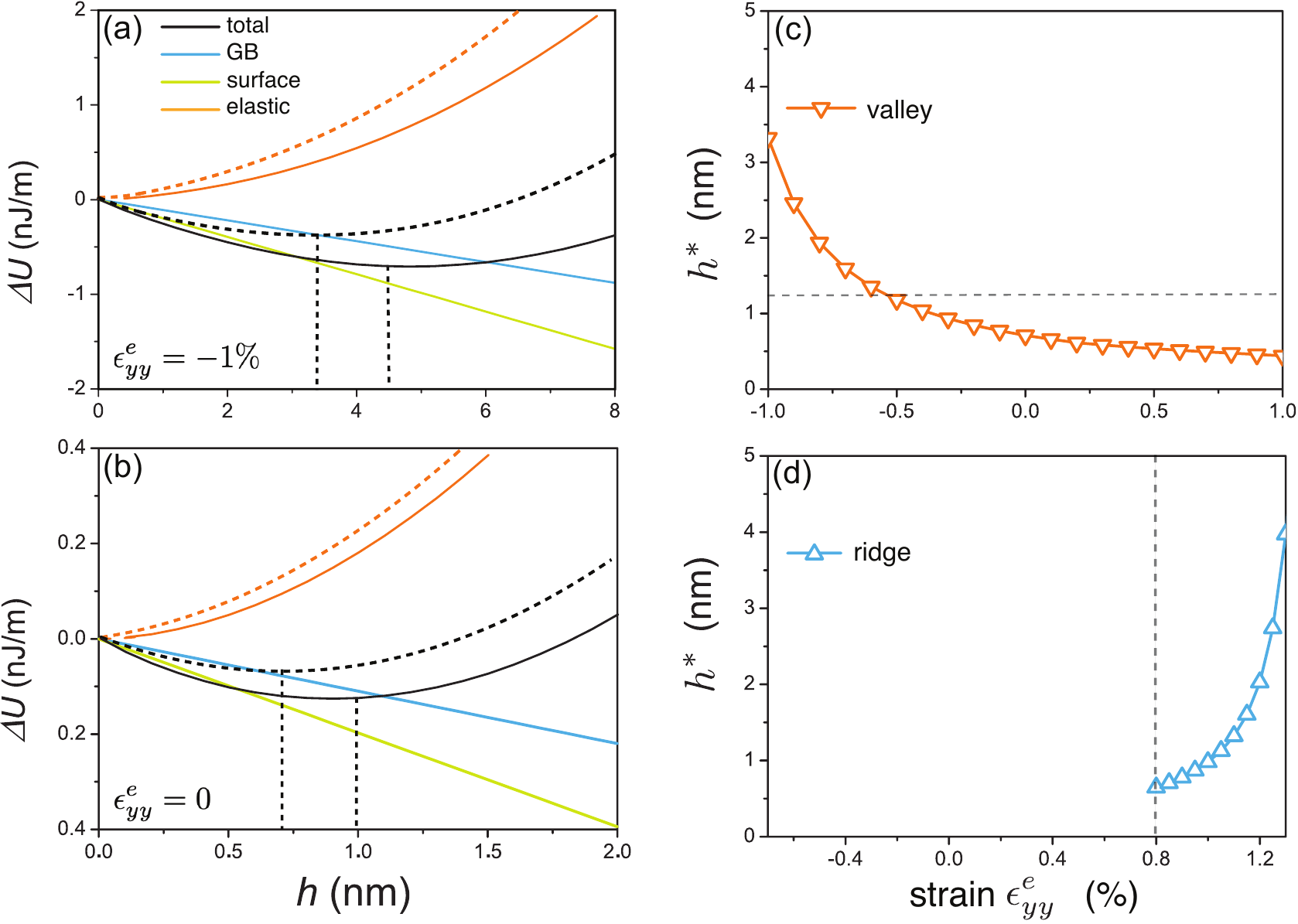}
\caption{The energy of the HAeGB system as a function of the thickness of the rotated layer $h$ (a) with external compressive strain $\epsilon_{yy}^e=-1\%$ and (b) in the absence of external strain $\epsilon_{yy}^e=0\%$. {In both cases, elastic energy curves without the surface elasticity contributions are also plotted (dashed lines)}. For compressive strain, the equilibrium thicknesses with and without the surface elasticity are $h^\ast=3.320$\,nm and $h^\ast=4.515$\,nm, respectively. In the absence of the external strain, the thickness values decrease to $h^\ast=0.712$\,nm and $h^\ast=1.000$\,nm, as indicated. (c-d) Equilibrium thickness of the {partially rotated layer $h^\ast$ for the HAeGB system} as a function of tensile and compressive applied strains $\epsilon_{yy}^e$ for (c) a valley and (d) a ridge based on the continuum framework. The dashed line in (c) is the thickness within a strain-free film $h^\ast(0)=1.250$\,nm extracted by energy minimization of the eGB system within all-atom MS computations~\cite{tsf:WangUpmanyuBoland}. The dashed line in (d) separates the stable and unstable regions for the ridge formation.
\label{fig:energyMin-hvsStrain}}
\end{figure*}


\subsubsection{Surface elastic energy cost $\Delta U_{e, s}$}
The surface is approximated as a 2D linear transverse orthotropic elastic material with a constitutive relation that satisfies Hooke's law~\cite{nano:Cammarata:1994, nw:MillerShenoy:2000}. The surface stress $\boldsymbol{\tau}(\tau_{xx},\tau_{yy})$ is
\begin{align}
\label{eq:HookesLawSurface}
& \tau_{xx} = \tau^0_{xx} + S_{xx} \epsilon_{xx} + S_{xy} \epsilon_{yy}, \\
& \tau_{yy} = \tau^0_{yy} + S_{xy} \epsilon_{xx} + S_{yy} \epsilon_{yy}\,, \nonumber
\end{align}
where $(\tau^0_{xx}, \tau^0_{yy})$ and $(S_{xx}, S_{xy}, S_{yy})$ are the relevant initial surface stress and surface elastic constants. 
The elastic response of the surface is absorbed within a composite framework consisting of bulk copper of thickness $H$ capped by two infinitesimally thin surface shells, as shown in Fig. \ref{fig:surfaceCompositeLayer}a. These shells deform in parallel with the bulk and hence they have the same in-plane deformation $\boldsymbol{\epsilon}(\epsilon_{xx}, \epsilon_{yy})$.
Then, the surface elastic energy is
\begin{align}
\label{eq:ElasticEnergySurface}
U_{s} =\,  & U - U_{b} \nonumber\\
	     = \, &  \tau^0_{xx}\epsilon_{xx}+\tau^0_{yy}\epsilon_{yy} \nonumber\\
	     & +\frac{1}{2}S_{xx}(\epsilon_{xx})^2+\frac{1}{2}S_{yy}(\epsilon_{yy})^2 + S_{xy}\epsilon_{xx}\epsilon_{yy} \, .
\end{align}
where $U$ is the total elastic energy (surface and the bulk) and $U_{b}$ is the bulk elastic energy devoid of free surfaces.


The surface elasticity parameters are extracted using MD simulations of the uniaxial and biaxial elastic deformation of $\langle111\rangle$ copper films of thickness $H=6.26$\,nm. The surface deformation energy $U_{s}=U - U_{b}$ is  plotted as a function of applied strain in Fig.~\ref{fig:surfaceCompositeLayer}b. Curve fits based on Eq.~\ref{eq:ElasticEnergySurface} yield the relevant parameters, $\tau_{xx}^0=\tau_{yy}^0=1.238$\,J/m$^2$, $S_{xx}=S_{yy}=1.807$\,J/m$^2$, and $S_{xy}=0.837$\,J/m$^2$. The positive or negative values of surface elastic constants indicate that the surface is stiffer or softer relative to the bulk.

The parameters are used quantify the elastic energy associated with the deformation of the surface within the eGB system, $\Delta U_{e,s}$. Copper is elastically anisotropic, and we simplify the in-plane strain deformation by assuming the bulk to be an isotropic material with modulus $E=128.5~\rm{GPa}$ and Poisson's ratio $\nu=0.345$~\cite{intpot:Mishin:2001, copper:Zhang:2016}.  For the $\langle111\rangle$ copper surface, the thickness of the surface layer is based on the cut-off distance for the nearest neighbor interactions ($0.55$\,nm). The reduction in the nearest neighbors within this surface layer leads to deviations from the bulk cohesive energy. We take the {surface layer thickness} to be $h_s=~0.626$\,nm, or thickness of 3 $\langle111\rangle$ layers. Then, the surface deformation modifies the energetics of eGB systems with equilibrium rotated layer thickness of the order of the surface layer, that is $h^\ast\sim h_s$. Based on prior estimates, we expect this to be the case for the HAeGB. On the other hand, for the $\theta=3.89^\circ$ LAGB, the equilibrium thickness of the rotated layer that { has been} previously extracted $h^\ast=15$\,nm is much higher than the surface layer thickness of $\langle111\rangle$ copper films~\cite{tsf:WangUpmanyuBoland}. The surface elasticity corrections are therefore negligible for {the low angle eGB (LAeGB) system}. 

The constitutive law of the linear orthotropic surface layer can be expressed in terms of the modulus $G$ and a new set of effective elastic constants $C_{ij}$ defined below:
\begin{align} \label{constitutive_law_new}
& \sigma_{xx} = \sigma^b_{xx} + \frac{\tau_{xx}}{h_s} = C_{11}(\epsilon_{xx}-\epsilon^0)+C_{12}\epsilon_{yy}+C_{13}\epsilon_{zz},  \nonumber\\
& \sigma_{yy} = \sigma^b_{yy} + \frac{\tau_{yy}}{h_s} = C_{12}\epsilon_{xx}+C_{11}(\epsilon_{yy}-\epsilon^0)+C_{13}\epsilon_{zz}, \nonumber\\
& \sigma_{zz} = \sigma^b_{zz} = C_{13}\epsilon_{xx}+C_{13}\epsilon_{yy}+C_{33}\epsilon_{zz}, \\
& \sigma_{xy} = G\epsilon_{xy}, \, \sigma_{yz} = G\epsilon_{yz}, \, \sigma_{zx} = G\epsilon_{zx}\,.\nonumber
\end{align}
The strain $\epsilon_0$ is the initial uniaxial contraction due to the surface stress within the orthotropic surface layer. It is size independent for film thicknesses $H\gg 2*h_s$. FEM computations using the {computational cell} shown in Fig.~\ref{fig:surfaceCompositeLayer}c yield the effective elastic constants $C_{11}=204.751$\,GPa, $C_{33}=201.865$\,GPa, $C_{12}=107.663$\,GPa, $C_{13}=106.326$\,GPa, $G=47.762$\,GPa, and initial surface strain $\epsilon^0=-0.877$\,\%.



For large grain sizes and thick films $h/L\ll1$ and $h/H\ll1$, these elastic constants together with Eqs. \ref{eq:surfEnergy} and~\ref{eq:bulkElasticEnergy2} can be used to quantify the bulk and surface elastic deformations, expressed as a function of $h_s/h$,
\begin{align}
\label{eq:totalElasticEnergy}
\Delta U_{e,b} + \Delta U_{e,s} = \, &\Gamma_s\left(\frac{h_s}{h}\right) \, G (\phi h)^2 \nonumber\\
& + \Omega_s \left(\frac{h_s}{h}\right)\, \epsilon_{yy}^e\, G \phi h^2
\end{align}
where the surface functions $\Gamma_s(h_s/h)$ and $\Omega_s(h_s/h)$ are 
\begin{equation}
\begin{aligned}
& \Gamma_s \left(\frac{h_s}{h}\right) =\left\{
\begin{aligned}
& \Gamma(0) \left[1 + A_s \frac{h_s}{h} + B_s (\frac{h_s}{h})^2 \right] \nonumber, \frac{h_s}{h}<1\\
& \Gamma(0) \left(1 + A_s + B_s \right) \nonumber, \frac{h_s}{h} \geq 1\\
\end{aligned}
\right.\\
& \Omega_s\left(\frac{h_s}{h}\right) = \Omega(0)\,.\nonumber
\end{aligned}
\end{equation}
As before for the corresponding bulk function, $\Omega_s$ does not exhibit any dependence on $h_s/h$ as the surface elastic constants are not affected by shear deformations. Combining Eqs.~\ref{eq:energyGB}, \ref{eq:surfEnergy} and \ref{eq:totalElasticEnergy}, the net energy change for the {restructured} eGB with a rotated layer of thickness $h$ (Eq.~\ref{eq:Energy1}) becomes
\begin{align}
\label{eq:energyTotal}
\Delta U = & G \phi \left[\Gamma_s\left(\frac{h_s}{h}\right) \phi + \Omega_s \left(\frac{h_s}{h}\right)\epsilon_{yy}^e \right] h^2 \\
& + \left[\frac{\gamma(\phi)}{\cos(\frac{\phi}{2})} - \gamma_{\langle111\rangle} \mp2\gamma_s \,h \tan\left(\frac{\phi}{2}\right)\right] h. \nonumber
\end{align}

The results of the FEM computations of the HAeGB system are summarized in Fig.~\ref{fig:surfaceCompositeLayer}d. The scaled elastic energy change $U_e(h_s/h, \epsilon_{yy}^e)/U_e(0, \epsilon_{yy}^e)$ is plotted with varying ratios $h_s/h$ and external strains $\epsilon_{yy}^e=0\%$, $-1\%$, and $-2\%$. Curve fits yield the relevant constants of the HAeGB system: $A_s=0.92, B_s=-0.45$ for a valley, and $A_s=-0.89, B_s=0.45$ for a ridge. Enhanced surface elasticity parametrized by $h_s/h$ leads to opposing trends for valleys and ridges, primarily due to the change in sign of the constant $A_s$. Since the surface stresses are tensile and valley formation entails tensile stresses within the surface layer, the elastic deformation energy increases and $h^\ast$ decreases. The opposite is true for ridge formation as it requires compressive stresses within the surface layer. The continuum framework based on Eq.~\ref{eq:totalElasticEnergy} yields similar results for the LAeGB. For the remainder of the article, we study the HAeGB as a representative eGB system.

\subsection{Equilibrium thickness of rotated layer $h^\ast (\phi, \epsilon_{yy}^e)$}
Figure~\ref{fig:energyMin-hvsStrain}a-b shows the  effect of the layer thickness $h$ on the changes in GB energy, surface energy and the elastic deformation energy for a valley formed at a {restructured} HAeGB in a copper film of thickness $H=50$\,nm. The layer thickness dependence of each of these terms is evident from Eq.~\ref{eq:energyTotal}. The curves correspond to a rotated layer with full reorientation of the misorientation axis towards a neighboring $\langle112\rangle$ direction, $\psi=\psi_{112}$. For the HAeGB, this corresponds to $\phi=\phi_{112}=9.097^\circ$.  As a measure of the effect of  surface elasticity, elastic energy curves with and without the surface deformation are plotted. The surface elasticity always decreases the overall elastic energy cost as it is softer compared to the bulk. For a compressive strains  $\epsilon_{yy}^e=-1\%$, the minimization of the total energy of the eGB system yields an equilibrium rotated layer thickness $h^\ast=3.320$\,nm. In the absence of surface elasticity, the value increases to $h^\ast=4.515$\,nm. Both minima are indicated graphically in Fig.~\ref{fig:energyMin-hvsStrain}a. While the difference is much smaller than the film thickness, the effect of surface elasticity is not insignificant, and we expect this contribution to become increasingly important at smaller film thicknesses.
\begin{figure*}[htp]
\includegraphics[width=1.8\columnwidth]{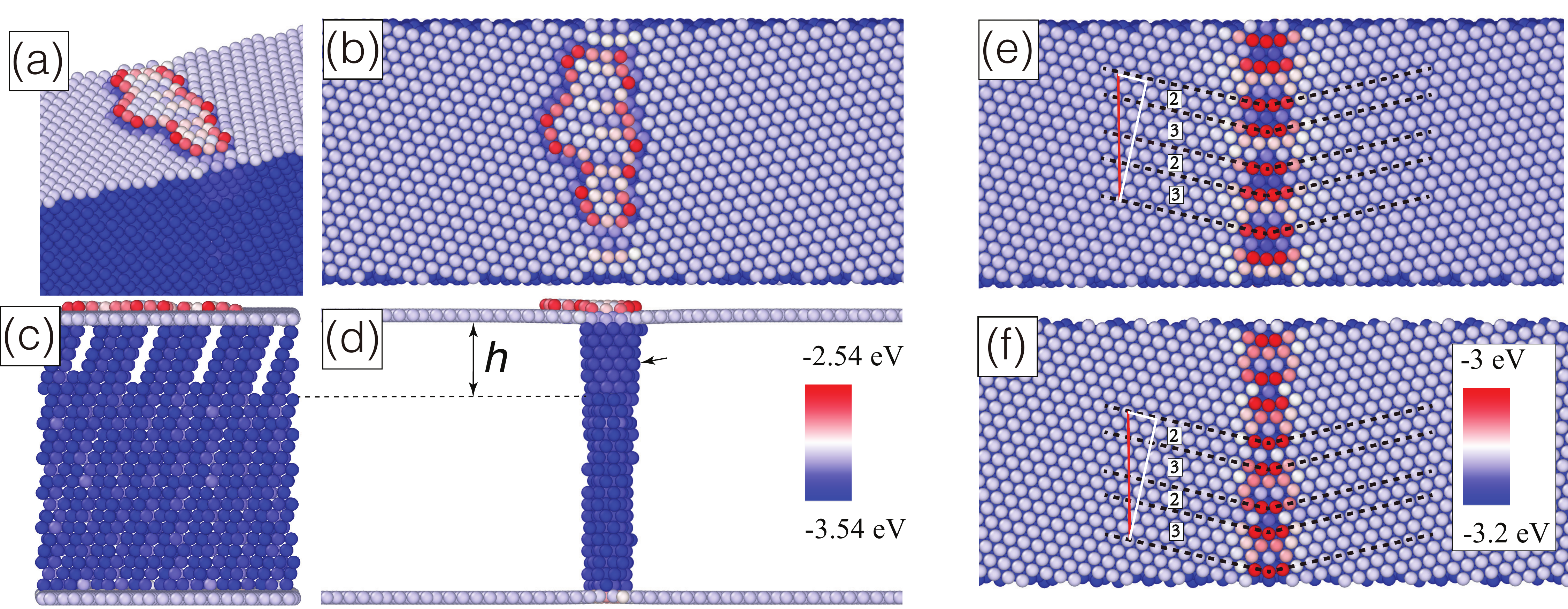}
\caption{Structural characterization of an initially flat eGB with a misorientation $+\theta$ within a $H=5$\,nm thick copper film compressively strained to $\epsilon_{yy}^e=-3\%$ at $T=700$\,K. (a) Perspective and (b) top views of the stepped island formation. (c) Side and (d) front views of sections through the GB core showing the GB { complexion} transition. The perfect FCC atoms are removed to clearly show the configurations of the GB atoms. (e-f) Surface TJ structure for the (e) compressively strained ($\epsilon_{yy}^e=-3\%$) and (f) strain-free $\epsilon_{yy}^e=0\%$ film. The $[3,2,3,2]$ decomposition of the periodicity vector (red solid line) is indicated. The surface island atoms in (e) are removed for visualization. Color scale represents the atomic potential energy.
\label{fig:atFig1-2}
}
\end{figure*}

In the absence of external strain, the minima decrease to $h^\ast=0.712$\,nm and $h^\ast=1.000$\,nm with and without surface elasticity. The corresponding plots are shown in Fig.~\ref{fig:energyMin-hvsStrain}b. Our recent atomic-scale computations on the HAeGB system using the CPSS framework have yielded an equilibrium rotated layer thickness of $h^\ast=1.25$\,nm~\cite{tsf:WangUpmanyuBoland}. Evidently, incorporation of the surface elasticity does not alleviate the disagreement between the continuum analysis and the atomic-scale computations, indicating that the disconnect is not due to surface elasticity based corrections. {Other contributing factors include the highly anisotropic elastic deformations in copper, the discrete nature of the atomic layers that make up the partially rotated layer, as well as the rotation induced elastic deformations that can modify the boundary structure and therefore their energies. These effects are absent within the continuum computations used to quantify the bulk and surface deformation energies. The continuum analysis also overestimates the contribution of the GB energetics to the driving force for the rotation as it is based on GB enthalpy changes; finite temperatures likely lower this driving force, necessitating a thicker rotated layer to balance the deformation energies. The comparison for layer thickness in strain-free films  also suggests that the continuum values serve as lower bounds for the equilibrium values $h^\ast$ in strained films.}

The strain dependence of the rotated layer thickness at valleys and ridges formed by the { restructured} HAeGB is plotted in Fig.~\ref{fig:energyMin-hvsStrain}c-d. The variation in the extrinsic strains is limited to $-1\% \le \epsilon_{yy}^e\le1.2\%$, well within the range where we expect the copper film deformations to be linear elastic. For a valley, compressive/tensile strains lead to increase/decrease in the layer thickness. That is, compressive strains promote {the grain rotation} at valleys while tensile strains inhibit the rotation. Additionally, the effect is non-linear at large strains. To see the basis for this trend, consider the limiting cases $h_s/h \ll 1$ and $h/H \ll 1$ where both surface deformation and film thickness based size corrections are negligible. Equation~\ref{eq:energyTotal} simplifies to
\begin{align}
\label{eq:energySimp}
\Delta U = & (0.153 + 12.133 \epsilon_{yy}^e) h^2 - 0.307 h~(\rm{valley}) \nonumber \\
\Delta U = & (0.141 - 11.072 \epsilon_{yy}^e) h^2 + 0.087 h~(\rm{ridge}). 
\end{align}
As expected, the energy change varies quadratically with the layer thickness $h$.  At a valley, for compressive strains greater than a small value $\epsilon_{yy}^e>-0.126\,\%$, the equilibrium thickness increases nonlinearly, 
\[
h^*\approx 0.154 / (0.153 + 12.133 \, \epsilon_{yy}^e).
\]
This is consistent with the trends in Fig. \ref{fig:energyMin-hvsStrain}c which also includes corrections due to surface elasticity contributions in a film of thickness $H=50$\,nm.  At a ridge, on the other hand, below a critical of $\epsilon_{yy}^e=0.127\,\%$ the ridge cannot form as the energy change is positive. Above this critical strain, the layer thickness increases non-linearly, again consistent with the trends shown in Fig.~\ref{fig:energyMin-hvsStrain}d. 

Our energetic analysis shows that the changes in the strain dependence of the partially rotated layer thickness at valleys  arise from the mechanical release of compression at the {restructured eGB} by efflux of atoms. Conversely, atom insertion at the valley relieves tensile strains, leading to decrease in the { thickness of the partially rotated layer}. For ridges, there is a critical tensile strain of $\approx0.8\%$ that is necessary to overcome the barrier associated with the increase in surface energy. This strain is much higher than the value expected in the limit $h\gg h_s$ (Eq.~\ref{eq:energySimp}), underscoring the importance of surface elasticity. Beyond this critical strain, the layer thickness increases rapidly with tensile strains as the stress at the eGB is relieved by atom influx.


 \begin{figure*}[htp]
\includegraphics[width=1.8\columnwidth]{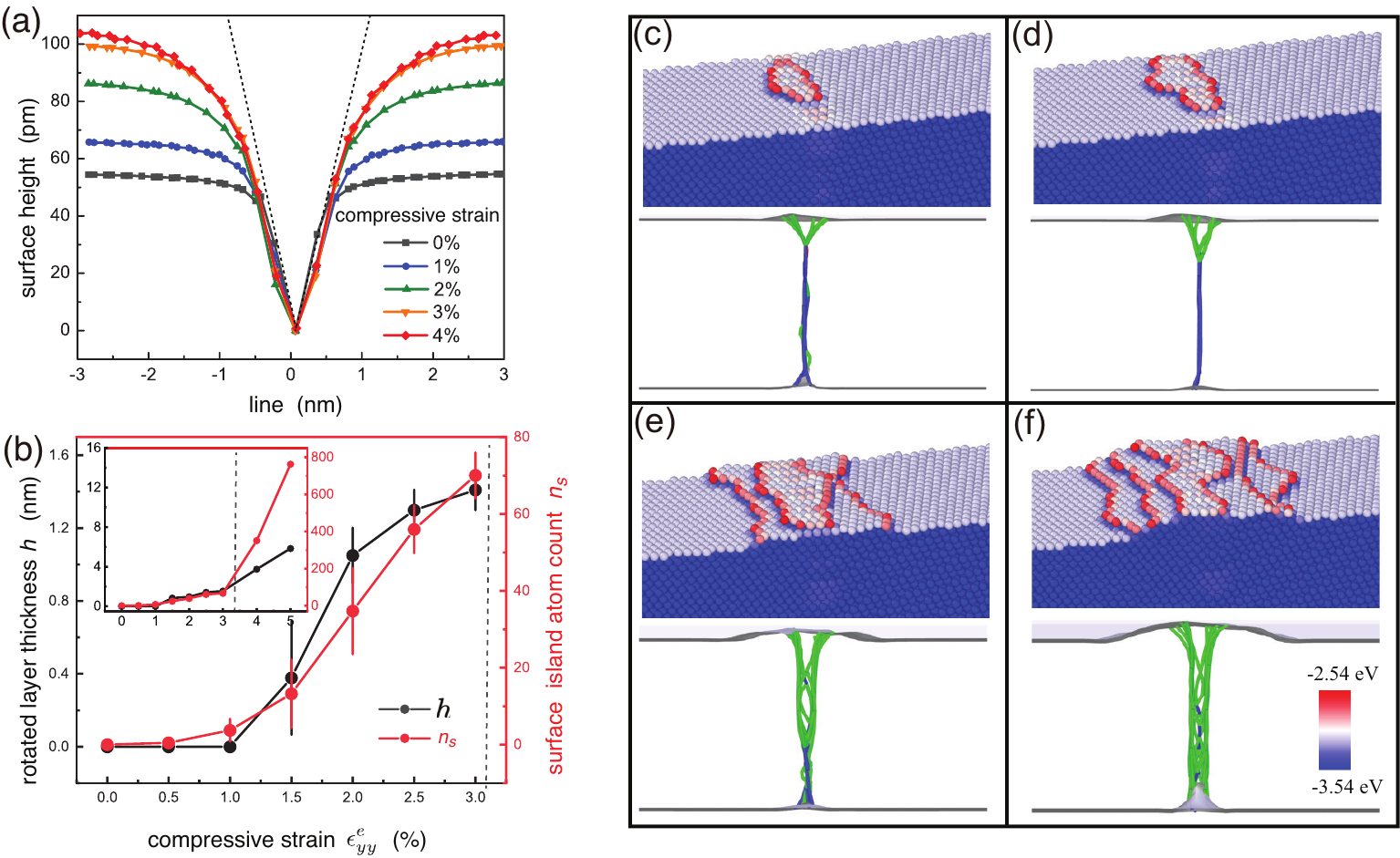}
\caption{(a) Compressive strain dependence of the average surface profiles across the surface TJ at $T=700$\,K. The error bars are of the order of symbol sizes and not shown for clarity.  Dashed lines indicate the expected local angle based on the out-of-plane rotation $\phi$. (b) Rotated layer thickness $h$ and number of island atoms $n_s$ (or island size) as functions of $\epsilon_{yy}^e$. (inset) The data $[h(\epsilon_{yy}^e), n_s(\epsilon_{yy}^e)]$ extracted from one set of simulations. The vertical dashed line in the vicinity of $-3\%$ indicates the onset of plastic deformation. (c-f) Step formation on top surface (top panels) and dislocation analysis of GB atoms (bottom panels) for compressive strains (c) $\epsilon_{yy}^e=-2\%$, (d) $-3\%$, (e) $-4\%$, and (f) $-5\%$. Color (from blue to red) indicates potential energy on top surface (from -3.54 to -2.54 eV). Blue and green colors indicate perfect dislocations and Shockley partials within the GB core, respectively.
\label{fig:atFig345}
}
\end{figure*}
\subsection{Atomic-scale simulations of the eGB system}
The continuum analyses show that for GBs with a local cusp in the vicinity of the growth direction, the thickness of the partially rotated top layer is sensitive to the sign and magnitude of the film stress. Here, we perform atomic-scale simulations of an initially flat (and therefore non-equilibrium) eGB subject to external film strains to study the formation of the rotated layer. We focus exclusively on the $\pm\theta=26.01^
\circ$ symmetric tilt HAeGB. The CPSS scheme that { Zhang {\it et al.}} have developed before to quantify the bulk deformation energy associated with the $\langle111\rangle\rightarrow\langle112\rangle$ rotation of the tilt axis~\cite{tsf:WangUpmanyuBoland}. We eschew this approach and use a combination of extrinsic stresses and temperatures that naturally arise during film growth to study the near-equilibrium response of the eGB. The approach makes contact with kinetic processes during film growth wherein the initial eGB formation is influenced by the substrate orientation and the transition can occur dynamically within the growing film. Systematic variations in the film strain are employed to study the nucleation and growth of the rotated layer and to identity mechanisms that stabilize the { eGB structural transition following grain rotation}.

\subsubsection{Strain-induced eGB structural transition: Valley}
We first study the response of a compressively strained eGB system in a copper film of height $H=5$\,nm. Simulations with larger film thicknesses show that this film thickness is sufficient to eliminate size effects based on the $\sim1$\,nm equilibrium rotated layer predicted by our continuum analysis. The initial eGB is composed of the bulk $\langle111\rangle - \theta=+26.01^\circ$ symmetric tilt HAGB terminating at a flat $\langle111\rangle$ surface. Relaxation of the system at $T=700$\,K in the absence of strain does not lead to an observable grain rotation, indicative of a barrier for the tilting of the misorientation axis of the HAGB. The lack of rotation persists to higher temperatures close to the bulk melting point (not shown). This is at odds with the $h^\ast=1.25$\,nm thick layer observed in atomic-scale computations based on the CPSS scheme~\cite{tsf:WangUpmanyuBoland}.

Compressive strains lead to a fundamentally different response. The surface configuration for $\epsilon_{yy}^e=-3\%$ at $T=700$\,K is shown in Fig.~\ref{fig:atFig1-2}a-b. We now see the nucleation and growth of an island along the surface TJ that equilibrates to a stable size. The surface displacements result in the formation of a valley. {Subsurface characterization of sections through the GB core shown in Fig.~\ref{fig:atFig1-2}c-d reveals a clear change in boundary structure to a finite depth}.  The GB dislocations within this layer reconstruct such that the stacking fault (SF) ribbons are oriented along $[111]$ planes (arrow, Fig.~\ref{fig:atFig1-2}d), consistent with the complete tilt of the misorientation axis towards the neighboring $[112]$ direction. The surface TJ structure underneath the island is depicted in Fig.~\ref{fig:atFig1-2}e. The atomic potential energy distribution shows dissociated SF ribbons at the surface, in contrast to the strain-free unrotated surface structure (Fig.~\ref{fig:atFig1-2}f). The surface TJ period vector reconstructs into a $[(3,1), (2,1), (3,1) , (2,1)]$ decomposition of the surface basis vectors $\frac{1}{4}[\bar{1}\bar{1}2]$ and $\frac{1}{4}[\bar{1}\bar{1}0]$, compactly referred to as the  $[3, 2, 3, 2]$ decomposition. It is identical to that observed for a strain-free and fully restructured eGB relaxed using the CPSS scheme~\cite{tsf:WangUpmanyuBoland}.

The surface profiles of the eGBs averaged over multiple lines across the surface TJ are plotted in Fig.~\ref{fig:atFig345}a. The equilibrium thickness of the layer averaged along the GB core is plotted in Fig.~\ref{fig:atFig345}b. In each case, the eGB is equilibrated at $T=700$\,K with the prescribed strain. For a strain-free (annealed) film, we see a groove form with a local angle $\approx9^\circ$, consistent with reconstruction of the surface TJ structure. The groove depth, defined respect to the flat surface far from the GB, is less than $55$\,pm. It quickly decays beyond the narrow GB core region, suggesting that there is an energy barrier for the eGB to fully restructure below the core of the surface TJ (Fig.~\ref{fig:atFig345}b). At $\epsilon_{yy} = -1\%$, the groove depth is $65$\,pm. We now observe the formation of a valley below the TJ within a partially rotated layer. The surface profile is characterized by a slow decay towards a planar orientation away from the eGB core (Fig.~\ref{fig:atFig345}a). The angle subtended at the eGB deviates from the local angle; we refer to this angle as the global angle. The eGB is fully restructured as the adjoining grains rotate out-of-plane. The valley formation is consistent with the $+\theta$ misorientation of the HAGB with respect to the vector along the axis of the added grain rotation $\Delta\hat{\rho}$ (Fig.~\ref{fig:schematicValleyRidge}b).   
                
The profiles in Fig.~\ref{fig:atFig345}a show an increase in the valley depths with compressive strains. The local angle at the groove remains the same, yet the global angle decays slowly to zero for strains larger in magnitude than $\epsilon_{yy}^e =-1\%$. Subsurface characterization shows that the {eGB does not restructure} for strains in the range $-1\%<\epsilon_{yy}^e\le0$. This is again in contrast to the theoretical calculations which predict a continuous increase in the equilibrium thickness to $h^\ast=3.32$\,nm for $\epsilon_{yy} = -1\%$. The valley depth increases modestly for strains in this range with a $[3, 2, 3, 2]$ reconstructed surface TJ.

Compressive strains beyond $\epsilon_{yy}=-1\%$ lead to a stable partially rotated layer with a spatially averaged thickness that increases monotonically with strain (Fig.~\ref{fig:atFig345}b). The strain sensitivity of the thickness decreases in the vicinity of $\epsilon_{yy}=-3\%$. At $\epsilon_{yy}=-4\%$ the layer thickness is $h=3.6$\,nm and the groove depth exceeds $100$\,pm. Comparisons with the continuum results for $h^\ast$ are no longer useful as the film is strained well beyond its linear elastic limit. Below $\epsilon_{yy}=-3\%$, the strain dependence $h(\epsilon_{yy}^e)$ is qualitatively similar to that from the continuum analysis (Fig.~\ref{fig:energyMin-hvsStrain}c-d).  The profiles exhibits significant departure from the local angle and the difference can be used to define a strain dependent global groove depth, 
\begin{align}
\label{eq:globalDepth}
d_g (\epsilon_{yy}^e) = \int_0^{\infty} [m(y) - m_0(y)] dy,
\end{align}
where $m(y)$ is the slope profile at a given distance $y$ from the strained eGB core and $m_0$ is the slope profile for a strain-free film. Both $d_g$ and  the rotated layer thickness $h$ arise due to the mechanical constraint from the $\langle111\rangle$ film away from the eGB core. The constraint is tuned by the external strain, and it is for this reason that we see similar trends in the strain dependence for both $d_g$ and $h$ in the simulations.  
\begin{figure}[htp]
\includegraphics[width=\columnwidth]{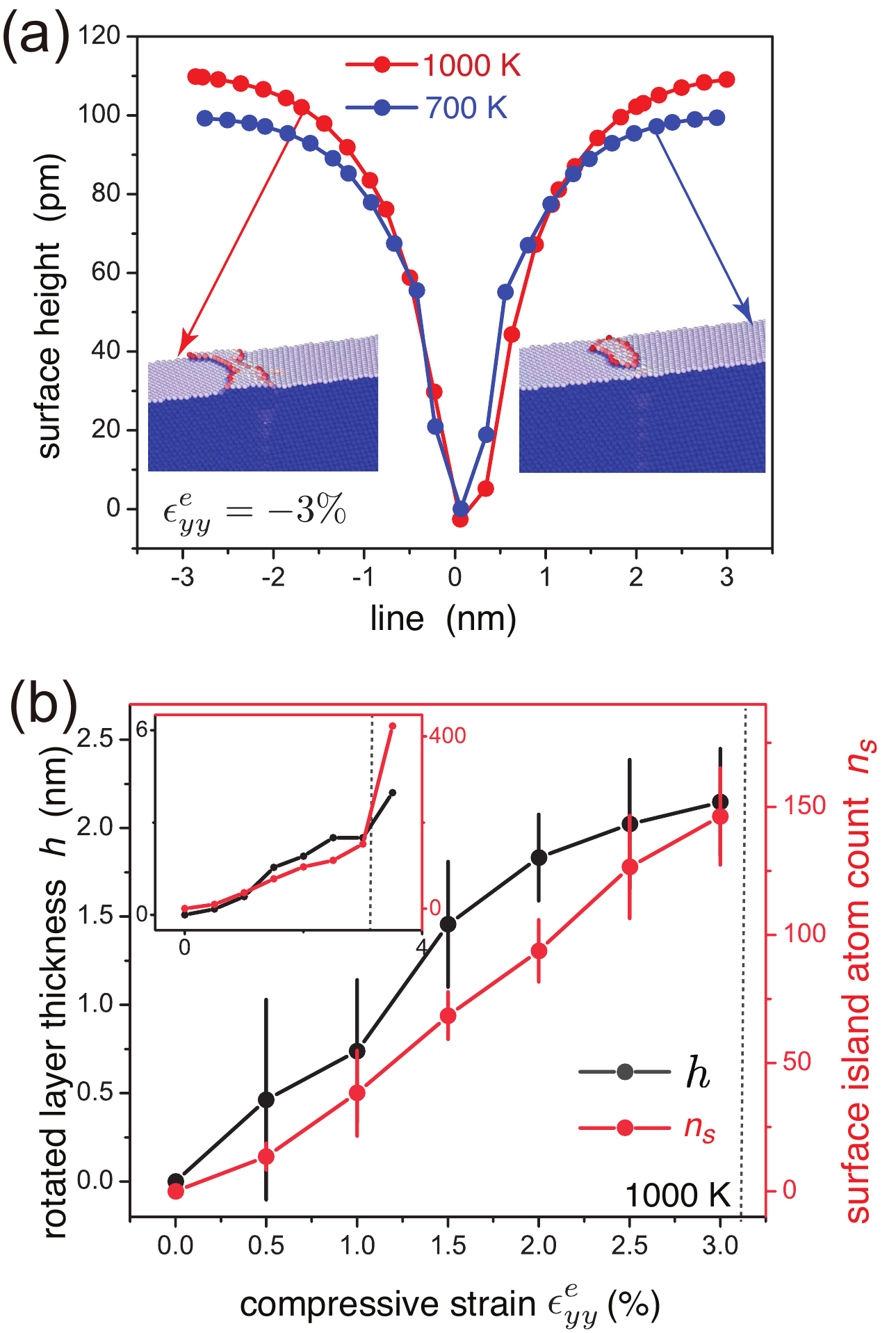}
\caption{(a-b) Same as in Fig.~\ref{fig:atFig345} but at $T=1000$\,K. (insets, a) Atomic configurations of the surface islands at $T=1000$\,K at $\epsilon_{yy}^e =  -3\%$. The configuration at $T=700$\,K is also shown for comparison. (inset, b) The thickness $h$ and island size $n_s$ based on one set of simulations with increasing compression. The vertical dashed line indicates the onset of plastic deformation {\it via} dislocation slip at the film surface.
\label{fig:atFig6-7}
}
\end{figure}


For each strain level that leads to a finite thickness of the partially rotated layer, the surface is decorated with surface islands. To quantify the extent of these islands, we monitor the number of atoms within the surface islands $n_s$ that stabilize on the {restructured} eGB system. The strain dependence is plotted in Fig.~\ref{fig:atFig345}b. The number of island atoms $n_s$ increases with the compressive strain, and the trend is similar to that for the thickness $h$. {The strong correlation between the two for small strains suggests that the formation of the rotated layer and the surface island are coupled. For strains larger than $\epsilon_{yy}^e=-2.5\%$, the strain sensitivity of $h$ decreases while the island atom count $n_s$ continues to increase (inset, Fig.~\ref{fig:atFig345}b), indicating the onset of another mechanism for increase in island size.}

To gain mechanistic insight, we track the configurations of the surface islands and dislocation structure of the subsurface GB atoms. Figure~\ref{fig:atFig345}c-f shows the island and GB configurations for strains larger than the critical value $\epsilon_{yy} \approx  -1\%$. Isolated surface islands form just above the critical strain in Fig.~\ref{fig:atFig345}c. The island size increases with strain, as expected based on the increase in $n_s$. Between $\epsilon_{yy}^e=-2.5\%$ and $\epsilon_{yy}^e=-3.0\%$, the islands begin to percolate into a new layer all along the surface TJ terminated by surface steps on either side {of the eGB} (Fig.~\ref{fig:atFig345}d). The rotated layer thickness increases, evident from the depth of the Shockley partials (green solid lines) at the edges of the $\{111\}$ SF ribbons. As we approach $\epsilon_{yy}^e = -4\%$, we observe a transition to a two-stepped island all along the TJ. {A representative} island configuration at $\epsilon_{yy}^e = -4\%$ is shown in Fig.~\ref{fig:atFig345}e. The leading edges of the enveloping step train form away from the eGB core, at the edges of the dissociated partials that terminate on the surface. The entire sequence is shown in Supplementary Movie 1. Examination of the surface configuration near the transition reveals that the new steps are formed following slip at the surface as the film surface starts to plastically deform. {The dislocations associated with the plastic deformation nucleate from the surface and are distinct from GB dislocations at the eGB core}. For the $\epsilon_{yy} =  -5\%$ configuration shown in Fig.~\ref{fig:atFig345}f, yet another layer forms as a result of the increasing strain accommodation via dislocation slip at the surface. Size effects also impact the eGB structure at these strain levels as the partially rotated layer thickness approaches the film thickness.
\begin{figure}[htp]
\includegraphics[width=\columnwidth]{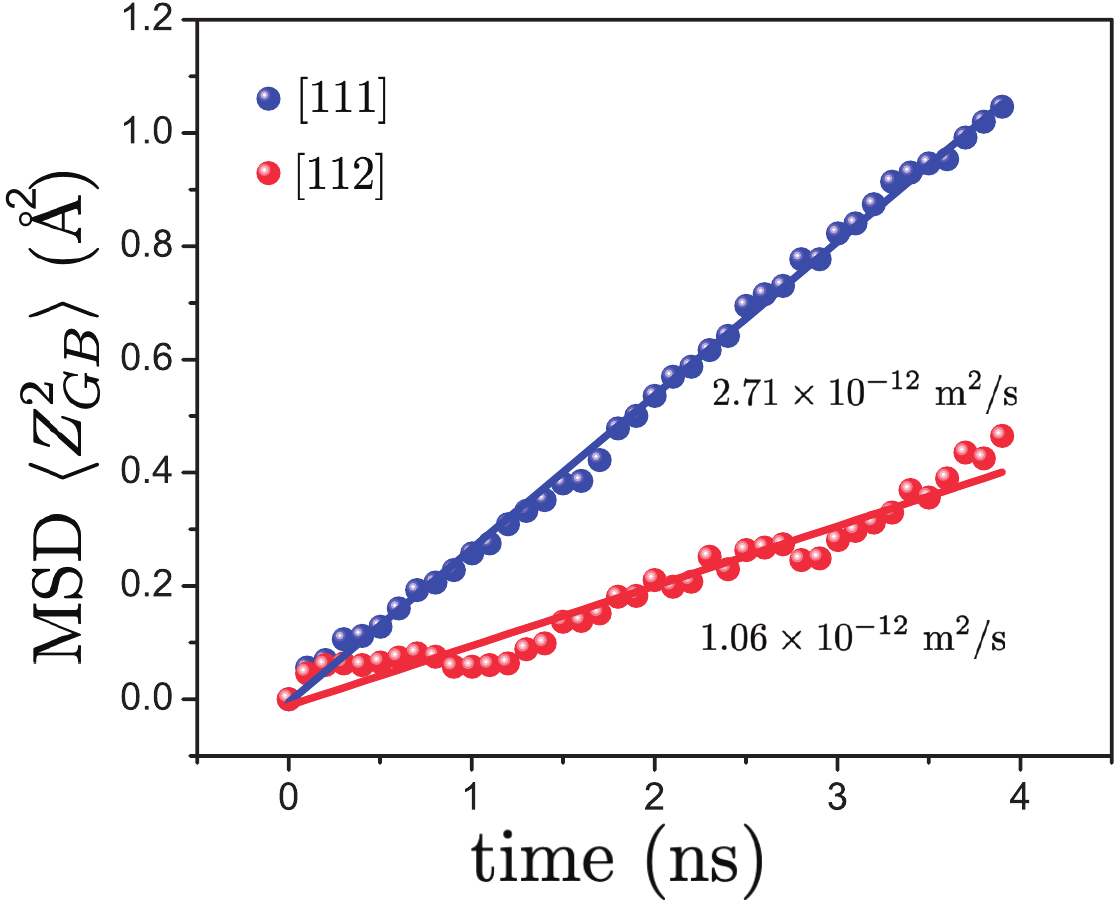}
\caption{Temporal evolution of the ensemble averaged mean squared atomic displacement (MSD) of {atoms within the GB core} $\langle Z_{GB}^2 \rangle$ at $T=1000$\,K for the rotated and unrotated GB. 
\label{fig:atFig8}
}
\end{figure}

Figure~\ref{fig:atFig6-7} summarizes the response at an elevated temperature, $T=1000$\,K. The $\epsilon_{yy}^e =  -3\%$ surface profile is plotted in Fig.~\ref{fig:atFig6-7}a. The local angle remains unchanged while the decay of the global angle is slower as the valley widens slightly compared to that at $T=700$\,K, or the global groove depth $d_g$ increases. The strain leads to the formation of surface islands, and comparisons with the configurations at $T=700$\,K show a clear increase in the island size (inset, Fig.~\ref{fig:atFig6-7}a). Isolated islands appear at $T=700$\,K, while at $T=1000$\,K they have percolated into an island layer around the { cores of the eGB dislocations}. Taken together, the trends in $d_g$ and surface island sizes suggest an increase in the rotated layer thickness with temperature. Strain dependence of the extracted average layer thickness $h$ as well as the number of island atoms $n_s$ plotted in Fig.~\ref{fig:atFig6-7}b confirm this to be the case. While the strain-free eGB is still unable to {restructure}, the critical strain for grain rotation is negligible. That is, the rotation is triggered as soon as a small compressive strain is applied. For each subsequent strain level, the partially rotated layer thickness is still smaller compared to the continuum predictions. 

The inset in Fig.~\ref{fig:atFig6-7}b reveals a relatively modest increase in $h$ between $\epsilon_{yy}^e =  -2.5\%$ and $\epsilon_{yy}^e =  -3\%$. The  increase in $n_s$ is still substantial in this range (not shown) as the GB dislocations slip. The critical strain for surface plastic deformation { in the vicinity of the eGB} is largely unchanged compared to that at $T=700$\,K. Beyond $\epsilon_{yy}^e =  -3\%$ the thickness and number of surface atoms increase rapidly and we see the formation of a second layer fueled by the slip of dislocations at the surface, similar to the configuration shown in Fig.~\ref{fig:atFig345}f.

The low strain response is clearly accelerated with temperature, suggestive of diffusive processes that serve as precursors to the formation of the rotated layer. Tracking the trajectories of the atoms during its formation reveals a finite diffusion flux primarily along the GB plane to the surface (not shown). To quantify this effect, we extract the diffusivity within {strain-free GBs before and after grain rotation} by monitoring the ensemble-averaged mean square displacements of the GB atoms $\langle Z^2_{\rm GB}\rangle$. Comparison of the MSDs of the two GBs at $1000$\,K shows that the diffusivity of the {GB after rotation} is smaller. This trend is true for temperatures in the range $600-1000$\,K (not shown) and is not surprising due to the higher density of the {restructured GBs} composed of dislocations oriented along the close packed $\langle111\rangle$ planes. It is then possible that the growth of the rotated layer after its nucleation slows down as the lower diffusion rates impede the incremental efflux of atoms from the eGB region necessary for the tilt of the misorientation axis. Compressive strains likely lower the diffusion rate further. 

\begin{figure}[htp]
\includegraphics[width=\columnwidth]{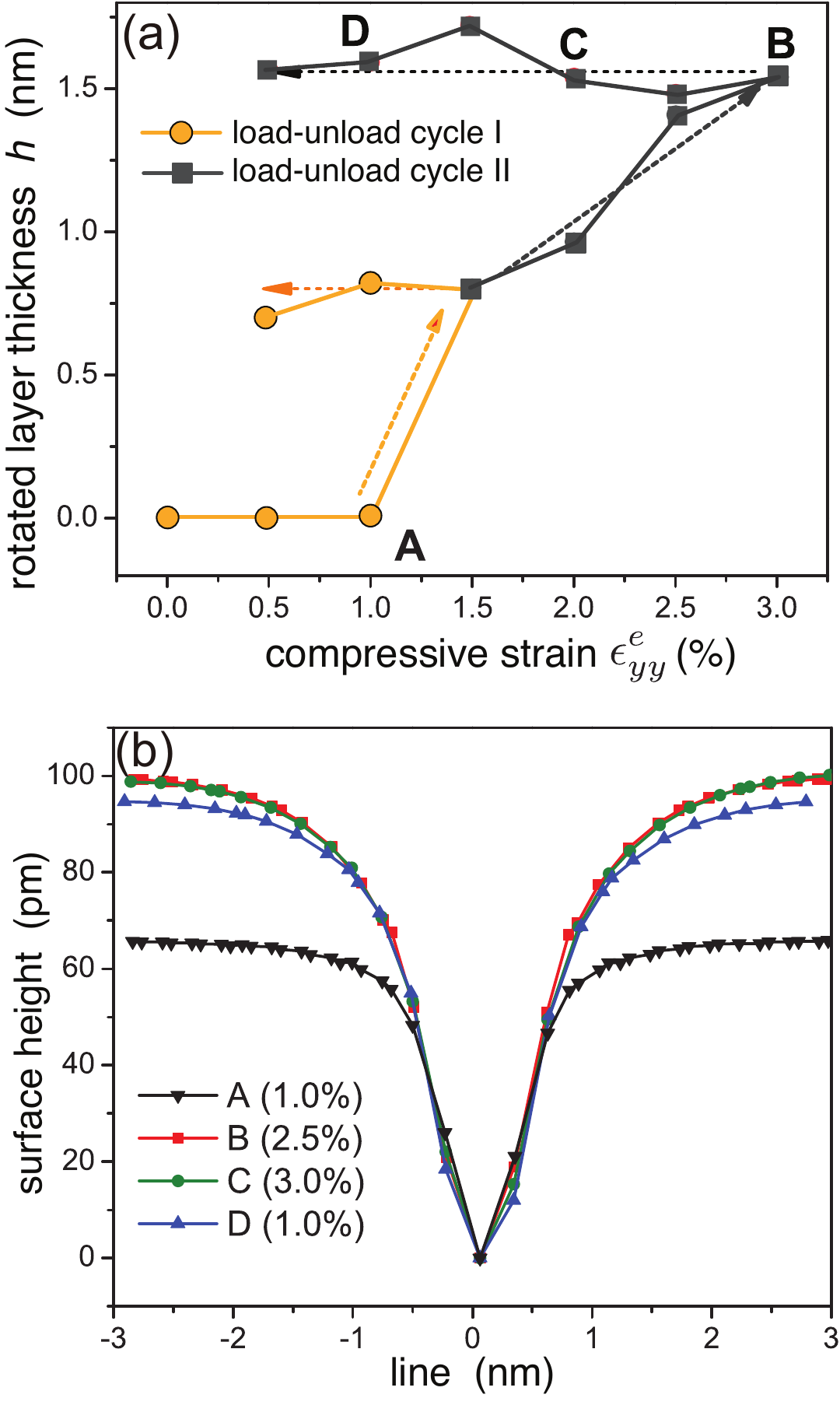}
\caption{(a) Strain dependence of the partially rotated layer thickness $h$ for two loading and unloading cycles (dashed yellow and black lines), as indicated. The solid lines indicate the layer thicknesses extracted from the simulations at intermediate values of the strain. (b) The surface profiles for the $\rm A\rightarrow B\rightarrow C\rightarrow D$ loading and unloading cycle. 
\label{fig:atFig9}
}
\end{figure}

Loading and unloading of the eGB bicrystal provides additional insight into the formation and stability of the partially rotated layer. The results for two loading-unloading cycles are plotted in Fig.~\ref{fig:atFig9}a. In both cases, an initial unrotated film with $\epsilon_{yy}^e = -1\%$ is chosen as the starting eGB structure (point A). The  surface profile plotted in Fig.~\ref{fig:atFig9}b reveals a shallow valley with a local angle that decays abruptly beyond the GB region, consistent with the $[3, 2, 3, 2]$ { reconstruction} of the surface TJ without the formation of an { elastically stressed rotated layer}. For the smaller loading cycle I, the film is loaded to $\epsilon_{yy}^e = -2\%$, allowed to equilibrate and then unloaded back to $\epsilon_{yy}^e = -1\%$. The rotated layer forms to a thickness of $h=0.8$\,nm at $\epsilon_{yy}^e = -2\%$ that co-exists with a stable surface island. On unloading to $\epsilon_{yy}^e = -1\%$, the layer thickness as well as the island size remain unchanged. While there is a small decrease at $\epsilon_{yy}^e = -0.5\%$ to $h=0.68$\,nm, the value is well below the equilibrium value for the  instrinsically reconstructed eGB.

The response for the larger cycle II $A\rightarrow B\rightarrow C \rightarrow D$  is similar. Loading to $\epsilon_{yy}^e = -3.0\%$ now increases the layer thickness to $h=1.55$\,nm (point B) with a larger island size. The thickness and the island size again remain largely unchanged on unloading to $\epsilon_{yy}^e = -2.5\%$ (point C) and then to $\epsilon_{yy}^e = -1\%$ (point D). { The surface profiles the valleys at $B$, $C$ and $D$ have depths in the range $90-100$\,pm}, convergent local angles, and global angles that decay slowly away from the GB region (FWHM~$\approx1.5$\,nm) with negligible strain dependence. The layer thickness is larger than the atomistic equilibrium thickness for strain-free eGB ($h^\ast=1.25$\,nm), and therefore unloading to zero strain should lead to a driving force for decrease in the layer thickness. Continued unloading to $\epsilon_{yy}^e = 0\%$ (not shown) has minimal effect on the layer thickness or the island size. The island size fluctuates but its average size remains unchanged in longer time-scale MD simulations, indicating that the thickness of the rotated layer is influenced by the compressive strain level and coupled to the energetic stability of island (and therefore its size) that persists to lower strain levels following unloading.   
\begin{figure*}[htp]
\includegraphics[width=1.8\columnwidth]{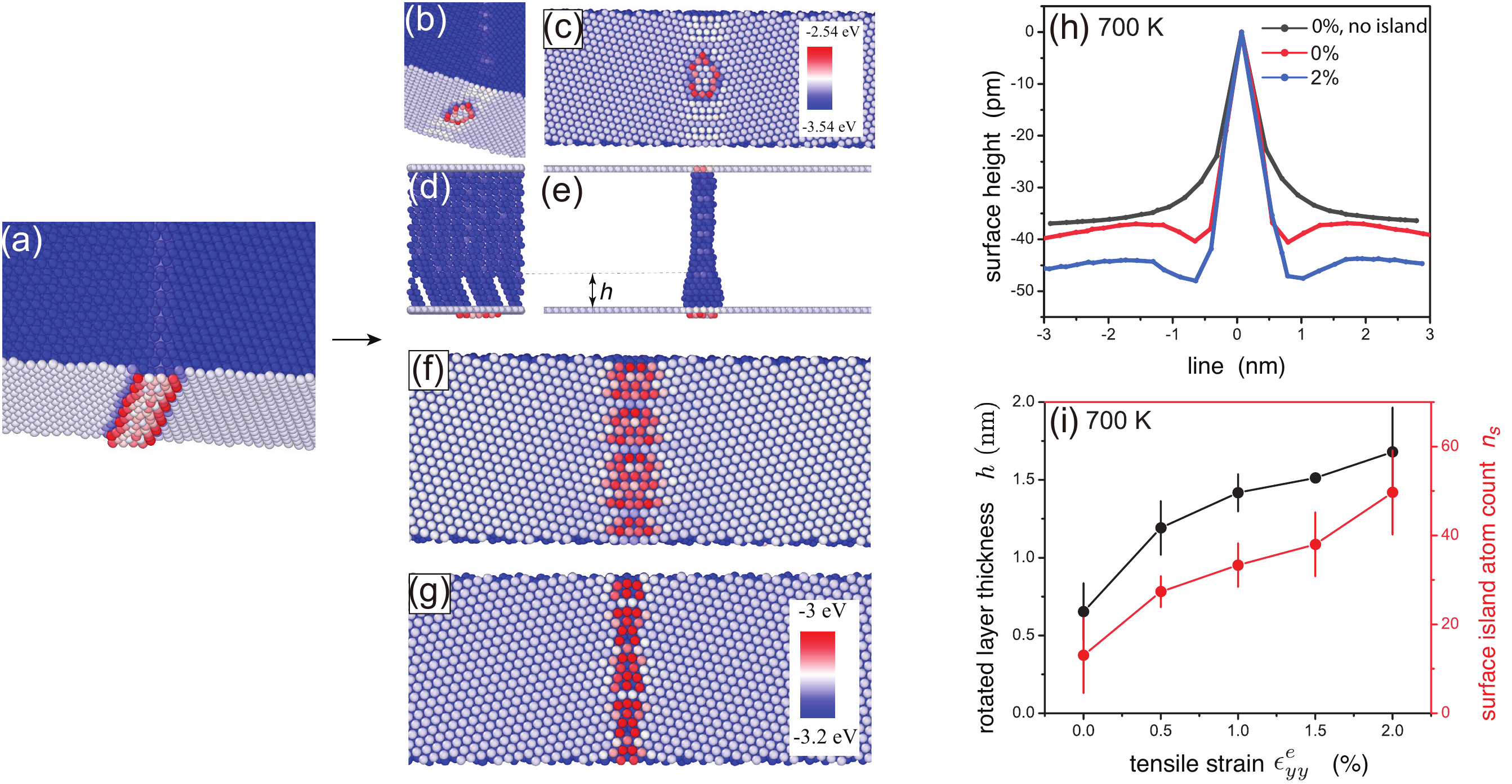}
\caption{(a) Initial configuration of an eGB with a 1nm wide surface island.  (b-g) eGB structure subject to 2\% tensile strain at $T=700$\,K. (b-e) Perspective, bottom, side, and front views of the surface island and the subsurface structure of the GB. The GB { complexion} transition can be observed in (d) and (e). (f) Surface TJ structure with tensile strain 2\% after removing the surface island atoms. The structure for the strain-free film is also shown for comparison in (g). (h) Surface line profiles across surface TJ in a $H=5$\,nm thick film for $\epsilon_{yy}^e=0\%$ and $\epsilon_{yy}^e=2\%$ at $T=700$\,K. The profile for the strain-free film without the surface island is also plotted for comparison. (i) The effect of the tensile strain on the layer thickness $h$ and the number of atoms absorbed by the GB from the surface island $n_s$ at $T=700$\,K. 
\label{fig:ridgeConfProfileThickness}
}
\end{figure*}

Atomic simulations of films under tensile strains do not lead to eGB valley formation. The rotation within the surface layer requires efflux of atoms from the eGB, while tensile strains applied along the GB normal can be mechanically relieved at the eGB by {\it influx} of atoms, thereby negating the driving force for the valley transition. This also implies that for a valley with an existing rotated layer, tensile strains act as a driving force for the layer to reduce its thickness, as seen in the continuum analysis (Fig.~\ref{fig:energyMin-hvsStrain}c). Changes in the layer thickness require an atomic influx towards (or vacancy efflux away from) the GB region. In the absence of such a source-sink action in the atomic-scale simulations, the tensile strains have no effect on eGB { complexion} formation or its growth.
\begin{figure*}[htp]
\includegraphics[width=1.8\columnwidth]{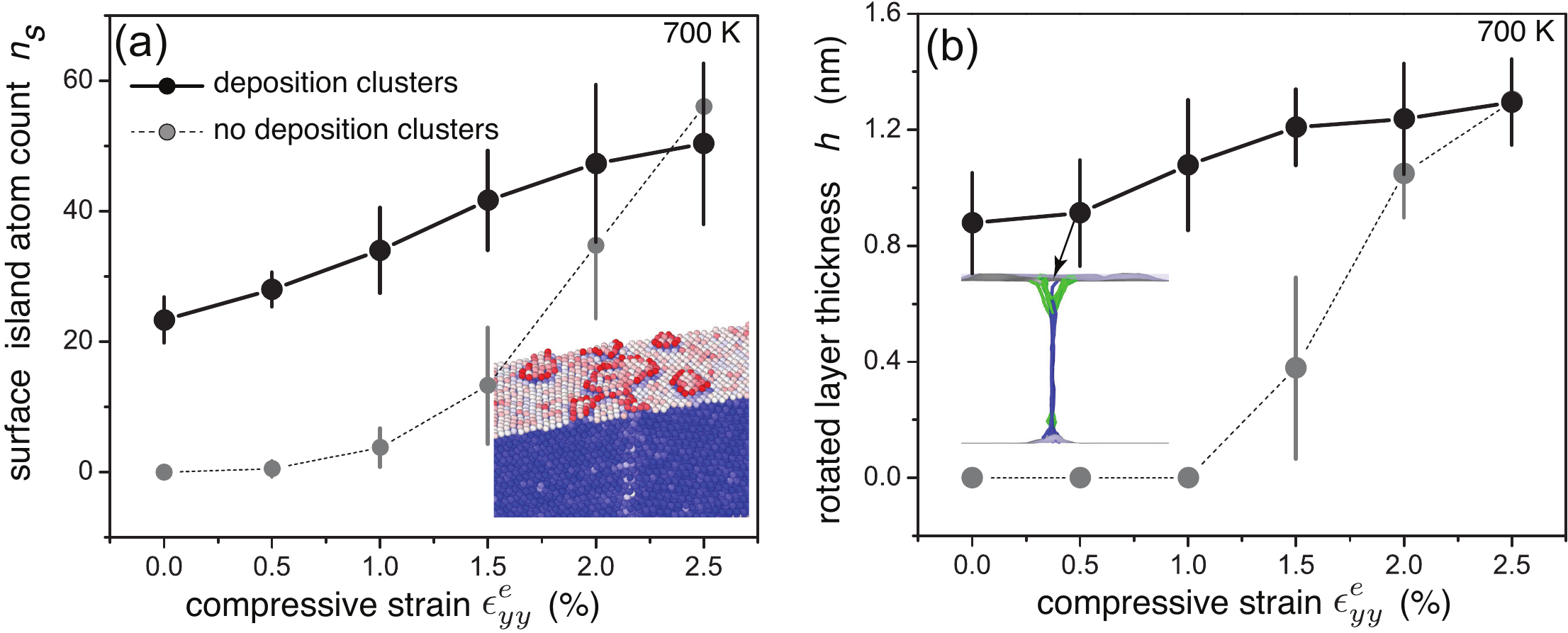}
\caption{(a) The effect of compressive strain $\epsilon_{yy}^e$ on (a) the number of surface island atoms and (b) the thickness of the rotated layer associated with valley formation $n_s$ in the presence of deposition clusters. The corresponding plots in the absence of these clusters are also shown for comparison (light gray dashed lines). (inset, a) Atom configuration showing the distribution of deposition clusters. (inset, b) Dislocation analysis of the subsurface structure at $\epsilon_{yy}^e=-0.5\%$ showing fully dissociated Shockley partials (green solid lines) within the rotated layer. 
\label{fig:atFig14}
}
\end{figure*}

\subsubsection{Strain-induced eGB structural transition: Ridge}
The theoretical framework shows that the formation of a ridge at an eGB with a $-\theta$  misorientation is facilitated by tensile strains. The surface area in the vicinity of the ridges increases and opposes the out-of-plane rotation. We then expect the tensile strains necessary for the structural transition to be larger than the magnitude of the compressive strains that led to the valley formation. Additionally, the ridge requires an atomic influx. In the absence of sources of atoms (or sinks for vacancies), nominal tensile and compressive strains cannot drive the structural transition of the eGB. Our simulations of eGBs confirm this to be the case.

{ To study the response in the presence of a source for atoms, we use initial configuration of the eGB with an equilibrated surface island strip of width $1$\, nm centered over the eGB, as shown in Fig.~\ref{fig:ridgeConfProfileThickness}a}. The configuration is analogous to the islands that stabilize over valleys. The $-\theta$ misorientated eGB is formed by the intersection of the HAGB in Fig.~\ref{fig:atFig1-2} with the bottom surface of the film. For compressively strained films, there is no transition at this eGB and the islands remains stable. On the other hand for both strain-free films and  films under tension, the strip width decreases, suggesting an influx of atoms into the eGB core.  
The atomic configurations following the GB structural transition at $\epsilon_{yy}^e=2\%$ are shown in Fig.~\ref{fig:ridgeConfProfileThickness}b-e. There is a marked decrease in the size of the surface monolayer and it pinches off into smaller surface islands. Subsurface GB structure consists of a restructured GB to a depth of $h=1.45$\,nm that is identical to that that forms below the compressively strained valleys, indicating a full rotation of the misorientation axis towards the $[112]$ orientation. The thickness decreases to $0.65\pm0.05$\,nm for strain-free films. We note that although the driving force for eGB valley formation is lower due to the contribution of surface energy to the transition embodied in~Eq.~\ref{eq:surfEnergy}, comparable compressive strains do not induce a transition at eGB valleys. The surface TJ exhibits the $[3, 2, 3, 2]$ decomposition at both $\epsilon_{yy}^e=0\%$ and $\epsilon_{yy}^e=2\%$ strain levels (Fig.~\ref{fig:ridgeConfProfileThickness}f-g). The width of the boundary is enhanced (FWHM $\approx 1$\,nm at $\epsilon_{yy}^e=2\%$) as it elastically deforms to accommodates the tensile strain, consistent with top layer rotation.  

Fig.~\ref{fig:ridgeConfProfileThickness}h shows the surface profiles below the surface island for $\epsilon_{yy}^e=0\%$ and $\epsilon_{yy}^e=2\%$ strain films. The $\epsilon_{yy}^e=0\%$ profile without the surface island is also plotted for comparison. {Clearly, the surface island strip drives the structural transition of the eGB and the ridge formation within the strain-free film. The eGB core between the rotated grains at the center of the ridge resides within a local groove}, a signature of the structural transition of the eGB and the formation of the rotated layer (Fig.~\ref{fig:figure1}e). The depth of the local groove changes from $\approx35$\,nm to  $\approx40$\,nm in the presence of the island strip. Tensile strain of $\epsilon_{yy}^e=2\%$ results in an increase in depth to $\approx60$\,nm.  The strain dependence of the thickness of the rotated layer is shown in Fig.~\ref{fig:ridgeConfProfileThickness}i. We see a monotonic increase in the thickness for strains in the range $0\le\epsilon_{yy}^e\le2\%$, indicating an absence of a critical strain for the eGB { complexion} nucleation. Beyond $\epsilon_{yy}^e=2\%$, the eGB deforms plastically via slip at the dissociated GB dislocations at the surface, similar to response for eGB valleys. Although the surface energy reduces the driving force for the { eGB restructuring and the the ridge formation} (Eq.~\ref{eq:surfEnergy}), the results indicate that the pre-existing surface island provides the atomic flux and effectively eliminates the barrier { for the structural transition}. The thickness of the rotated layer is still smaller than the continuum analysis, quite like the trends observed for valleys.    
\begin{figure*}[htp]
\includegraphics[width=1.8\columnwidth]{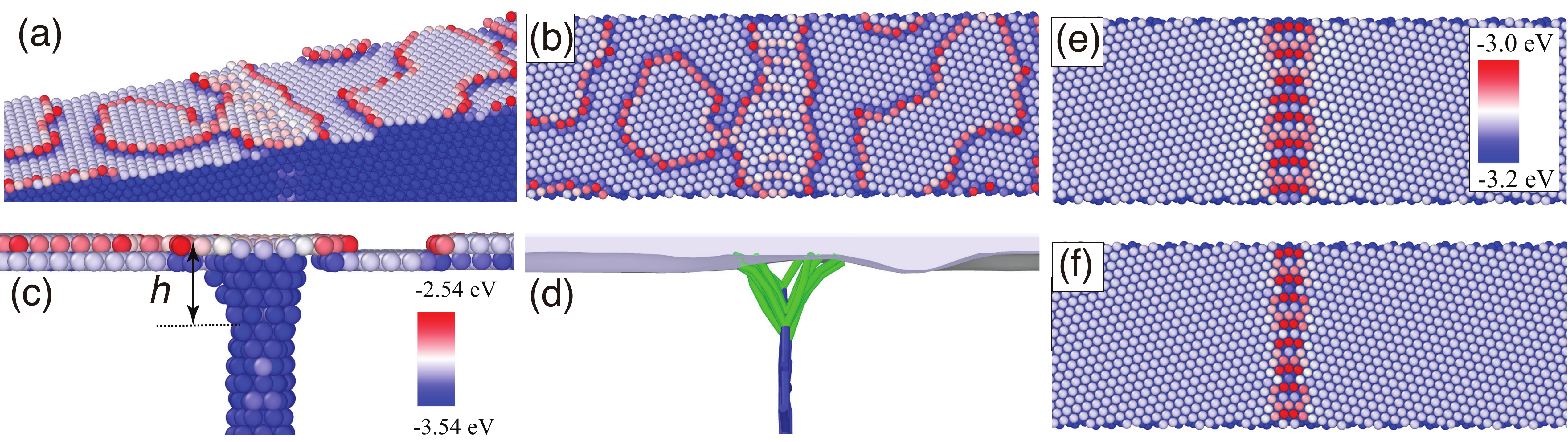}
\caption{Structure analysis for deposition-induced GB transition at $T=700$\, K. The deposition rate is fixed at 0.1 atoms/ps. (a-d) 3D, bottom, side, and front views on top surface due to additional atomic extrusion to step edges. The GB { complexion} transition within finite thickness can be observed in (c) common neighbor and (d) dislocation analyses.  (e-f) Triple junction structure of the eGB { within the strain-free film} (e) with and (f) without the deposition flux. Color scale indicates potential energy.
\label{fig:atFig15}
}
\end{figure*}

\subsubsection{Deposition induced eGB structural transition: Valley} 
{ The strain-induced restructuring of the eGBs reveal that} the island nucleation and the driving force for rotation $\Delta U $ are not decoupled. To better understand their role in the transition, we perform thin film deposition simulations where the adatom flux naturally allows the surface islands to form and also provides sources and sinks for atomic flux to or away from the eGB. The deposition rate, deposition velocity and temperature are fixed at $0.1$\,atom/ps, $100$\,m/s, and $700$\,K respectively, allowing for significant relaxation via adatom diffusion between deposition events.  


We first study the effect of small clusters during early stages of the deposition on a strain-free $+\theta$ HAeGB. One such atomic configuration is shown as an inset in Fig.~\ref{fig:atFig14}a. Small isolated yet mobile clusters form on the surface. The configuration corresponds to an earlier stage of film growth~\cite{tsf:GrabowGilmer:1980} compared to that used to study the formation of the ridge (Fig.~\ref{fig:ridgeConfProfileThickness}a). Agglomeration of these clusters at the { cores of the eGB dislocations} leads to the formation of a stable (critical) surface island that consists of the deposited as well as initial film atoms. Evidently, there is a finite atomic flux away from the eGB towards the surface that aids in the formation and growth of the surface clusters. Supplementary Movie 2 shows the mechanistic details of the island formation. The substructure characterization shows the formation of an $h=0.85\pm0.15$\,nm thick rotated layer. 

The application of compressive strains accelerates the valley formation. The surface island atoms $n_s$ and layer thickness $h$ plotted in Fig.~\ref{fig:atFig14} rapidly increase with strain. At $\epsilon_{yy}^e=0.5\,\%$, the layer has grown to a thickness of $h=0.9\pm 0.15$\,nm. Dislocation analysis of the substructure reveals that the Shockley partials within the rotated layer are { dissociated} (inset, Fig.~\ref{fig:atFig14}b); as before, the GB axis is fully rotated towards the $[112]$ axis. It is notable that the effect of compressive strains in the absence of the deposition clusters, also plotted for comparison in Fig.~\ref{fig:atFig14}, is different in that the eGB restructures only under the application of a critical compressive strain. Evidently, the barrier for the { nucleation of the eGB complexion and the partially rotated layer} is removed by the presence of the surface clusters. 

Figure~\ref{fig:atFig15}a-b shows the atomic configurations of the strain-free $+\theta$ HAeGB after $t=8$\,ns, or the deposition of $800$\,atoms. The simulations are performed at $T=700$\,K to facilitate layer-by-layer growth of the islands. We see the islands on the surface enveloped by meandering steps. Subsurface structural characterization over multiple simulations show the presence of a $h=1.04$\,nm thick rotated layer (Fig.~\ref{fig:atFig15}c-d). It is comparable to that formed by the deposition clusters during early stages of the growth and is closer to the equilibrium value of $h^\ast=1.25$\,nm. The surface TJ structure under the deposited layers (Fig.~\ref{fig:atFig15}c-d) is identical to that formed in the absence of the deposition flux. Atom insertion into the eGB can lead to compressive stresses that in turn can drive the eGB transition. However, at this deposition temperature, the atomic flux is directed outward along the GB towards the surface and it is facilitated by mobile clusters that move into the eGB core and trigger the valley formation. The simulations show that eGB { complexion} formation is influenced by the size and morphology of the surface islands that nucleate and grow on the surface during deposition. 

Supplementary Figure 1 shows the eGB structure 
at a higher deposition rate of $1$\,atom/ps and lower temperature, $T=300$\,K. Smaller islands form on the growing layer before it completely forms. { The departure from layer-by-layer growth is due to the combination of high deposition flux and low deposition temperature that limits the adatom diffusivity}.  The eGB again undergoes the structural transition and the rotated layer thickness is similar to that at slower deposition rates, $h\approx 1$\,nm. The lower temperature increases the kinetic barrier for the growth of the partially rotated layer due to the reduced GB diffusivity, and therefore amplifies the effect of the island nucleation and growth on the eGB structural transition. The robustness of the rotated layer thickness to a completely different island morphology indicates that deposition-induced formation and growth of the layer thickness occurs mainly during early stages of growth. Thereafter, it settles to a steady-state that is determined largely by the stress evolution within the growing film.

{\section{Conclusions}}
Our continuum framework and atomic-scale simulations capture salient aspects of the effect of film strains on the co-existing GB { complexions} formed at symmetric tilt eGBs in $\langle111\rangle$ copper films. Our main findings are summarized below:
\begin{enumerate}
\item
The equilibrium thickness of the rotated surface layer $h^\ast$ formed by the $[111]\rightarrow[112]$ tilt of the misorientation axis at the eGB is sensitive to the strain in the film. The { grain rotation towards the lower energy GB complexion} results in valleys at eGBs with positive misorientations while ridges are favored at eGBs with negative misorientations. Unlike valleys, ridge formation requires a critical tensile strain as the change in the surface energy reduces the driving force for the ridge transition. 
\item
Compressive strains increase while tensile strains decrease the thickness of the rotated layers at eGB valleys. The opposite is true at ridges. This interplay between strain and $h^\ast$ stems from the stress relief at the eGBs which modifies the balance between the stored elastic deformation and the changes in the interfacial (GB and surface) energies associated with the transition. 
\item
Surface elasticity based corrections depend on the relative thicknesses of the surface layer $h_s$, the rotated layer $h^\ast$, and the film thickness $H$.  It is increasingly important for thin films with $h^\ast$ of the order of the surface layer, $h^\ast \sim H$ and $h^\ast\sim h_s$. 
\item
The surface profiles at the eGBs reflect the interplay between the external strains and the rotated layer thickness. Local angles form {at the eGB core}, consistent with the expected out-of-plane rotation {due to the $[111]\rightarrow[112]$ tilt of the misorientation axis}. Compressive strains widen the valleys with a slower decay of the profile away from the eGB core and a concomitant increase in $h^\ast$. An analogous effect occurs at ridges in response to tensile strains. The strain dependence is characterized by changes in the global groove depths $d_g$, a parameter that quantifies the departure from the local groove angles due to the mechanical constraints imposed by the substrate and the film strain (Eq.~\ref{eq:globalDepth}). As such, both $h^\ast$ and the global groove depth $d_g$ are local measures of the film strain. 
\item
The formation of eGB valleys and ridges that relieve the external strains requires atomic efflux/influx from the eGB to atom (or vacancy) sources and sinks. In the absence of these fluxes, the eGB is unable to undergo any structural transition. The (in-)efficiency of the diffusive pathways can lead to kinetic barriers for the eGB {structural transition}, compounded by the low diffusivity of the compact GB { complexions} that form following the transition. Low intrinsic mobility of the GB interphase line defect can also contribute to the sluggish kinetics. These trends may in part explain the temperature dependent rotated layer thickness observed in the atomistic simulations. 
\item
Loading and unloading of the eGBs reveal a strain-dependent plasticity-like response wherein the surface island formation and growth occurs irreversibly with changes in the rotated layer thickness. Simulations with pre-existing islands as well as deposition fluxes confirm the presence of an energetic barrier coupled with the size of the surface island.     
\end{enumerate}


{\section{Discussion}}
{ Our conclusions are rooted in our energetic framework for the {stability of eGB complexions following grain rotation within stressed films}. The external film stresses modify the intrinsic strains generated in the vicinity of eGB due to the  grain rotation into a low energy  GB complexions~\cite{tsf:WangUpmanyuBoland}. The resultant elastic stress distribution modifies the extent of eGB restructuring and the surface morphology. We emphasize that the resultant surface morphology has less to do with activation volume changes between the rotated and unrotated GB { complexions}. They follow from the geometric accommodation of the rotation within the surface layer. The stability of this layer is determined by changes in the surface and deformation energies that lead to nucleation and growth of the lower energy GB { complexion}, suitably aided by diffusive fluxes to/from the eGB.

The results have far reaching implications for film growth. For suitably oriented substrates, the eGB structural transition into partially rotated surface layers can occur during the Volmer-Webber growth of polycrystalline films. The rotation is likely aided by the migration of the GBs as part of the evolution of the strained GB microstructure~\cite{grot:HarrisKing:1998, grot:Upmanyu:2006}. Our atomic-scale simulations show that the surface TJ at the eGB reconstructs and develops a local angle before the restructuring of the eGB. The transition occurs spontaneously in $T=0$\,K molecular statics simulations of strain-free eGBs, or it is almost activationless. The observation is consistent with the small change in energy of the surface TJ~\cite{tsf:WangUpmanyuBoland}. In effect, the reconstructed TJ 
serves as a heterogeneous site for the nucleation of the GB { complexion}. However, the eGB structural transition is kinetically arrested as it requires flux of atoms to or away from the eGB from (near-equilibrium) sources and sinks of atoms and vacancies such as surfaces, grain boundaries and related crystalline interfaces, and dislocations~\cite{book:SuttonBalluffi:1995}. Fast diffusion along the GB to the surface is a readily available pathway. This is evident from the simulations with and without pre-existing islands on the film surface which establish a strong coupling between formation of the surface islands and the eGB valleys and ridges.    
 
The eGB restructuring is likely to influence the coalescence of crystallites~\cite{tsf:Hoffmann:1976, tsf:NixClemens:1999} during early stages of film growth. The tensile stresses that develop favor the formation of ridges at $-\theta$ eGBs. However, the eGBs are strongly influenced by the mechanical constraint imposed by the substrate orientation. 
As an example, for high modulus substrates with intrinsic resistance to deformation, our energetic framework shows that the transition can be energetically prohibitive as the elastic deformation and excess surface energy necessary for the rotation cannot be balanced by the reduction in GB energy. The constraint, parameterized by the size correction $\Gamma(h/H)$, decreases as the strained film grows in thickness. Beyond a critical thickness we therefore expect an eGB structural transition into a ridge,
aided by accelerated influx of atoms. Relaxation via atom insertion at the eGBs during subsequent growth~\cite{tsf:PaoChasonSrolovitz:2002} leads to a reversal to compressive stresses.
Then, we expect a valley to form at the $+\theta$ eGBs. 
Additionally, the existing ridges adjust their layer depth and morphology in response to the decrease in tension. In both instances, atom efflux to the surface is necessary. 


For the coupled island-eGB system, the equilibrium layer thickness $h^\ast$ is one that co-exists with an infinitely large island, or a fully formed surface layer. It then follows that the finite island sizes at eGBs modify the chemical potential of atoms in the restructured eGB, and the equilibrium shifts to a different thickness of the rotated layer. As validation, for both strain-free and strained eGBs, the layer thickness formed dynamically in the atomistic simulations is smaller then the equilibrium value calculated using CPSS-based continuum analysis with surface elasticity based corrections. Since the continuum values are lower bounds, the thickness within the island-eGB coupled system is also lower than the $h^\ast$ extracted using CPSS-based atomistic simulations~\cite{tsf:WangUpmanyuBoland}. At $T=700$\,K, the eGB valley formation requires a critical strain (Fig.~\ref{fig:atFig345}b). Although the diffusivity of the rotated eGB is lower, it is still sufficient for fast diffusion of atoms to the surface to accommodate the necessary atomic efflux for the valley formation. As such, this critical strain is a measure of the energetic barrier related to the nucleation of a surface island. The dislocations {at the core eGB core} can lower this nucleation barrier, quite like spiral crystal growth in the presence of surface dislocations~\cite{tsf:BurtonCabreraFrank:1951}.

The coupled island-eGB framework offers insight into the effect of temperature and loading-unloading cycles. Increasing temperature reduces the critical strain for eGB valley formation, evident from the $h^\ast(\epsilon_{yy}^e)$ plots in Fig.~\ref{fig:atFig6-7}b. The island-eGB coupling is modified, likely due to changes in the energetics of the steps~\cite{cg:JeongWilliams:1999}.
The temperature effectively alters the chemical potential of the island-eGB system. We expect similar trends at ridges in the presence of atom sources. 

The changes in chemical potential of the island-eGB system also affect the structural evolution during unloading of the strained eGB valleys. During loading, the size of the supercritical islands increases with the external strain at the eGB valley as they form a new surface layer. For island strips, the monopole strength at each of the two oppositely signed steps is proportional to the product of the film stress and step height and it therefore increases~\cite{tsf:KourisSieradzki:1999}. At a given strain-level, the monopole-monopole attraction (together with the areal contribution due to difference in the stresses in the island and the strained surface below) sets the critical island island size. Unloading to a lower strain reduces the monopole-monopole attraction, further stabilizing the island to lower strain levels. While it is likely that the areal contribution also decreases, the atomistic simulations suggest that the former is the dominant effect, resulting in plasticity-like response wherein the islands (and therefore $h$) increase in size with incremental film strains and remain stable on unloading.

While it is tempting to generalize the strain-mediated {eGB structural formation} and stability to polycrystalline surfaces, this study is limited to a flat GB terminating at the $\langle111\rangle$ surface of an FCC copper film. It therefore represents the unconstrained response of the eGB system in a specific material system. GBs within columnar microstructures in thin films are typically associated with an in-plane curvature that changes the orientation of the GB plane, and this requires a careful consideration of the response of $+\theta$ and $-\theta$ {\it asymmetric} tilt eGBs. Since the $\langle112\rangle$ tilt axis no longer resides on the GB plane, the GB can undergo faceting transitions to maximize occupancy of close packed planes~\cite{gbe:HsiehBalluffi:1989, gbe:MuschikFinnis:1993, gbe:BargRabkin:1995}, or rotate to altogether different cusps. If there are no local cusps or the GB itself resides at a local cusp (e.g. coherent twin boundaries), it may not {restructure} at all. The structure at each eGB can also be influenced by other eGBs in the vicinity of surface quadrijunctions where they meet. Several of these aspects need to be resolved in order to fully realize the benefits of eGB engineered surfaces. In the emerging world of emergent grain boundaries and interfaces, there is plenty of room at the top!}\\
\\
\noindent
{\bf Funding}: This work was completed in part using the Discovery cluster, supported by Northeastern University?s Research Computing team. The authors are also grateful for supercomputing resources available at Northeastern University through the Massachusetts Green High Performance Computing Center (MGHPCC).  M.U. was supported in part by Army Research Office [grant numbers W911NF-12-R-0012 and W911NF1520026-CLIN0006]. M.W., R.Y., XH and H.W. acknowledge support from the National Natural Science Foundation of China (Grant No. 12172347) and the Fundamental Research Funds for the Central Universities (Grant No. WK2480000006).\\

\end{document}